\documentclass[nofootinbib,showkeys]{revtex4}
\usepackage{todonotes}
\usepackage{graphicx}
\usepackage{amsmath,amsthm,amssymb,bm,amsfonts}
\usepackage{empheq}

\usepackage{color}
\usepackage[makeroom]{cancel}
\usepackage[normalem]{ulem}
\definecolor{kgbcolor}{rgb}{0,0.1,0.7}
\definecolor{ascolor}{rgb}{1,0,1}

\newcommand\kgbout{\marginpar{\color{kgbcolor}$\clubsuit$}\bgroup\markoverwith{\color{kgbcolor}{\rule[0.4ex]{2pt}{0.8pt}}}\ULon}
\newcommand\asout{\marginpar{\color{ascolor}$\heartsuit$}\bgroup\markoverwith{\color{ascolor}{\rule[0.4ex]{2pt}{0.8pt}}}\ULon}

\def\Phat{{P}}

\newcommand{\tdm}[1]{\mbox{\boldmath$#1$}}

\def\aprime{a^\prime}

\def\tprime{t^\prime}
\def\abis{a^{\prime\prime}}

\def\ftilde{\tilde{f}}

\def\Atilde{\tilde{A}}
\def\Etilde{\tilde{E}}
\def\Ptilde{\tilde{P}}
\def\Dtilde{\tilde{D}}

\def\Ehat{\hat{E}}
\def\Phat{\hat{P}}
\def\Dhat{\hat{D}}

\def\kperp{k_\perp}
\def\pperp{p_\perp}

\def\rperpb{\mathbf{r}_{\perp}}
\def\kperpone{k_{1 \perp}}
\def\kperptwo{k_{2 \perp}}
\def\kperponeb{\mathbf{k}_{1 \perp}}
\def\kperptwob{\mathbf{k}_{2 \perp}}

\def\kperpb{\mathbf{k}_{\perp}}
\def\kappaperp{\kappa_\perp}
\def\kappaperpi{\tdm{\kappa}_{i \perp}}
\def\kperpthree{k_{3 \perp}}
\def\kperpthreeb{\mathbf{k}_{3 \perp}}

\newcommand{\eq}{\!&=&\!}

\newcommand{\be}{\begin{eqnarray}}
\newcommand{\ee}{\end{eqnarray}}

\begin{document}
\title{\bf Unintegrated double parton distributions}

\author{Krzysztof Golec-Biernat}
\affiliation{Institute of Nuclear Physics, Polish Academy of Sciences, 31-342 Cracow, Poland}
\affiliation{Faculty of Mathematics and Natural Sciences, University of Rzesz\'ow,  35-959 Rzesz\'ow, Poland}

\author{Anna M. Sta\'sto}
\affiliation{Department of Physics, The Pennsylvania State University, University Park, PA 16802, United States}

\begin{abstract}
We present the  construction of unintegrated double parton distribution functions  which include dependence on   transverse momenta
of partons. We extend the  formulation which was used to obtain  the single unintegrated parton distributions from the standard, integrated parton distribution functions. Starting from the homogeneous part of the  evolution equations for the integrated double parton distributions,  we construct the   unintegrated double parton distributions  as the  
convolutions of the integrated double distributions and the  splitting functions, multiplied by the Sudakov form factors. We show that there exist three  domains of external hard scales which require three distinct forms of the unintegrated double distributions. 
The additional transverse momentum dependence which arises through the Sudakov form factors leads to non-trivial correlations in the parton momenta.
We also discuss the non-homogeneous contribution to the unintegrated double parton distributions,   which arises due  to the  splitting  of a single parton into daughter partons with high transverse momenta.
We analyze two cases, the unfolding of the transverse momenta dependence from the last step of the  evolution of two partons, and the case where the transverse momenta are generated directly from  the single parton  splitting.
\end{abstract}


\maketitle

\section{Introduction}

The Large Hadron Collider opened a completely new kinematic domain for exploring the dynamics of the strong interactions. At these very high energies the incoming hadrons are characterized by the large parton densities driven by the fast increase of the gluon density at low values of Bjorken $x$. Typically, for most hadron encounters only a single partonic interaction occurs. However, at large energies, it is also possible to have more than one partonic interaction per one collision of incoming hadrons.  This is referred to as a multi-parton interaction. Such events were   first observed by the AFS Colaboration at CERN \cite{Akesson:1986iv}
and followed by the measurements performed by the collaborations at the Tevatron collider \cite{Abe:1997bp,Abe:1997xk,Abazov:2009gc}. 
Later, a systematic experimental analysis was performed at the  Large Hadron Collider \cite{Aad:2013bjm,Chatrchyan:2013xxa,Aad:2014rua}.

The theoretical description of single hard  parton interactions is well established within the perturbative QCD.
The standard approach is to use the collinear factorization \cite{Collins:1985ue,Collins:1989gx} with  perturbatively calculable partonic cross sections and integrated parton distribution functions (PDFs) which obey the  Dokshitzer-Gribov-Lipatov-Altarelli-Parisi (DGLAP) evolution equations \cite{Dokshitzer:1977sg,Gribov:1972ri,Altarelli:1977zs}. Such  factorization is well defined when the hard scale, like the transverse energy of the jet, invariant mass of the Drell-Yan pair or the mass of the produced heavy quark, is present in the process.

For the multiparton interactions, the theoretical description  within the perturbative QCD is also possible in the presence of the sufficiently hard scales.  The computation of double parton scattering (DPS)
cross sections within the collinear framework makes use of the double parton distribution functions (DPDFs) \cite{Shelest:1982dg,Zinovev:1982be,Ellis:1982cd,Bukhvostov:1985rn,
Snigirev:2003cq,Korotkikh:2004bz,Gaunt:2009re,Blok:2010ge,Ceccopieri:2010kg,Diehl:2011tt,
Gaunt:2011xd,Ryskin:2011kk,Bartels:2011qi,Blok:2011bu,Diehl:2011yj,Luszczak:2011zp,
Manohar:2012jr,Ryskin:2012qx,Gaunt:2012dd,Blok:2013bpa,
vanHameren:2014ava, Maciula:2014pla, Snigirev:2014eua,  Golec-Biernat:2014nsa, Gaunt:2014rua, 
Harland-Lang:2014efa, Blok:2014rza, Maciula:2015vza}.
Recently, a significant progress has been made towards a complete proof of the factorization theorem for the double parton interaction in the case of the double Drell-Yan production \cite{Diehl:2015bca}.
In the  leading logarithmic approximation, the  DPDFs obey QCD evolution equations  similar to the DGLAP  equations for the PDFs \cite{Kirschner:1979im,Shelest:1982dg,Zinovev:1982be,Snigirev:2003cq,Korotkikh:2004bz,Gaunt:2011xd},  (see also \cite{Konishi:1978yx,Konishi:1979cb} for the analogous formulation of the evolution equations for double parton correlations inside jets). 

The standard collinear approach with integrated PDFs may be, however, insufficient when trying to describe more exclusive processes, see for example \cite{Collins:2005uv}. In this case the more complete information about the kinematics of the partonic process should be included. This can be done by using the unintegrated parton distributions which in addition to the parton longitudinal momentum fractions also include  their transverse momentum dependence. 

The unintegrated parton distribution functions (UPDFs)\footnote{They belong to a general class of transverse momentum dependent parton distributions (TMDs).} naturally appear in the small $x$ formalism, where  the so called $k_T$-factorization is utilized \cite{Catani:1990eg} with off-shell matrix elements and the unintegrated parton distributions. For example, the Balitsky-Fadin-Kuraev-Lipatov (BFKL) equation \cite{Balitsky:1978ic,Kuraev:1977fs,Fadin:1975cb} can be interpreted as an evolution equation in $\log x$ for the unintegrated parton distributions.  
The Catani-Ciafaloni-Fiorani-Marchesini (CCFM) equation \cite{Ciafaloni:1987ur,Catani:1989sg,Catani:1989yc,Marchesini:1994wr} is a further example of the evolution equation for the UPDFs, which in addition to the transverse momentum also depends on the hard scale of the process. Yet another formulation is the transverse momentum dependent (TMD) factorization, (for a comprehensive formulation see  \cite{Collins:2011zzd}), which is valid to the leading power in the hard scale.

A very useful approach to the UPDFs
was formulated in \cite{Kimber:1999xc,Kimber:2000bg,Kimber:2001sc}, where the  UPDFs  were constructed from the integrated PDFs through the derivative of the latter additionally  dressed with the Sudakov form factor. 
The inclusion of the Sudakov form factor  leads to the emergence of the dependence on two scales, the transverse momentum of the parton and the hard scale.  The hard scale plays the role of the  cutoff in the angular ordering of the emitted soft gluons. This construction is relatively convenient as it allows for obtaining the UPDFs without actually solving
separate equations (like the CCFM equation which is quite complicated) but rather using the standard  integrated PDFs.
The UPDFs obtained in this framework are widely used in phenomenology, where they are applied in the  $k_T$ factorization formalism together with the  off-shell matrix elements, see \cite{Kutak:2016ukc} and references therein for recent analysis.

In this article, we extend the construction \cite{Kimber:1999xc,Kimber:2001sc} to the case of the  unintegrated double parton distribution functions
(UDPDFs)\footnote{They are also called double transverse momentum dependent distributions (DTMDs) in the current literature, see recent \cite{Buffing:2016qql}.}. Starting from the evolution equations for the integrated double distribution functions (DPDFs) we recast them in the form that allows to extract the unintegrated versions of these distributions. We show that for the homogeneous part of the solution to these equations, the extension requires the convolution of the integrated DPDFs with splitting functions and multiplication by the appropriate Sudakov form factors. Since there are two hard scales in this case, we find that the form of the UDPDFs depends on the relation between the two hard scales.
Also, we find that the cutoffs which regularize the real emission integrals and the Sudakov form factors induce nontrivial correlations between the longitudinal momenta of the two partons. 

We also discuss  the  non-homogenous contribution to the UDPDFs which corresponds to the splitting of one parton into two daughter partons with large transverse momenta. 
We present the results of  the unfolding of the transverse momenta dependence from the last step in the  evolution of two partons in the non-homogeneous part of the solution 
to the evolution equations for the DPDFs. We also consider the  contribution  due to the parton splitting which includes transverse momentum dependence,  derived in the light-front approach. 

Our paper is organized as follows. In Sec.~\ref{sec:skmr}  we recapitulate the construction of the UPDFs presented in 
\cite{Kimber:1999xc,Kimber:2001sc}   in  both  the Mellin space and the $x$-space.  
In Sec.~\ref{sec:dpdf} we recall the evolution equations for the integrated DPDFs and also show their formulation in the Mellin space. In Sec.~\ref{sec:udpdf} we present the details of the construction of the UDPDFs for the homogeneous part of the solution the evolution equations for the DPDFs.
We also briefly discuss  the correlations between kinematic variables in the UDPDFs, induced by the regularization of the real emission integrals and the Sudakov form factors. In Sec.~\ref{sec:nonhom} we construct the non-homogeneous contribution to the UDPDFs, by first applying the construction performed for the homogeneous solution and also by the explicit derivation of the parton splitting term with the transverse momenta dependence on the light-front. Finally, in the last section we present the summary and conclusions.

\section{Unintegrated parton distributions}
\label{sec:skmr}

Let us first recapitulate the main points of the construction of the single {\it unintegrated parton distribution functions} 
proposed in  \cite{Kimber:1999xc,Kimber:2001sc}.
The starting point are the DGLAP evolution equations for the single integrated parton distributions $D_a(x,\mu)$, where
$a$ denotes quark/antiquark flavors and also gluon, $x$ is the longitudinal momentum fraction and $\mu$ is the  scale for this distribution. The DGLAP equations with real and virtual parts separated read
\be
\label{eq:1}
\frac{\partial{D_a(x,\mu)}}{\partial \ln \mu^2} = \sum_{\aprime}\int_x^{1-\Delta} \frac{dz}{z}\,
P_{a\aprime}(z,\mu)\, D_{\aprime}\Big(\frac{x}{z},\mu\Big)
- D_a(x,\mu)\,\sum_{\aprime}\int_0^{1-\Delta} dz z P_{\aprime a}(z,\mu)\,.
\ee
The splitting functions $P_{a\aprime}$ can be computed order by order in perturbation theory and thus are given in powers of  the running strong coupling constant, $\alpha_s(\mu)/(2\pi)$. In the 
 leading logarithmic in $\mu^2$  approximation we have
\be
\label{eq:1a}
P_{a\aprime}(z,\mu) = \frac{\alpha_s(\mu)}{2\pi}\,P^{(0)}_{a\aprime}(z)\,,
\ee
where $P^{(0)}_{a\aprime}$ are the LO Altarelli-Parisi splitting functions.
The upper limits
in the divergent integrals in Eq.~(\ref{eq:1}) are regularized by a parameter $\Delta<1$ to be specified later. In the DGLAP equations $\Delta\to 0$ because singularities between real and virtual terms  cancel, but we will keep $\Delta$ finite to be able to manipulate these equations.
The first (real) term on the r.h.s of Eq.~\eqref{eq:1} can be interpreted as a number of partons which are emitted in the interval  $\mu^2 \le \kperp^2 \le\mu^2+\delta \mu^2$. The second (virtual) term does not change the transverse momentum of the parton and therefore can be integrated as we shall show below.  

\begin{figure}
\begin{center}
\includegraphics[width=0.30\textwidth]{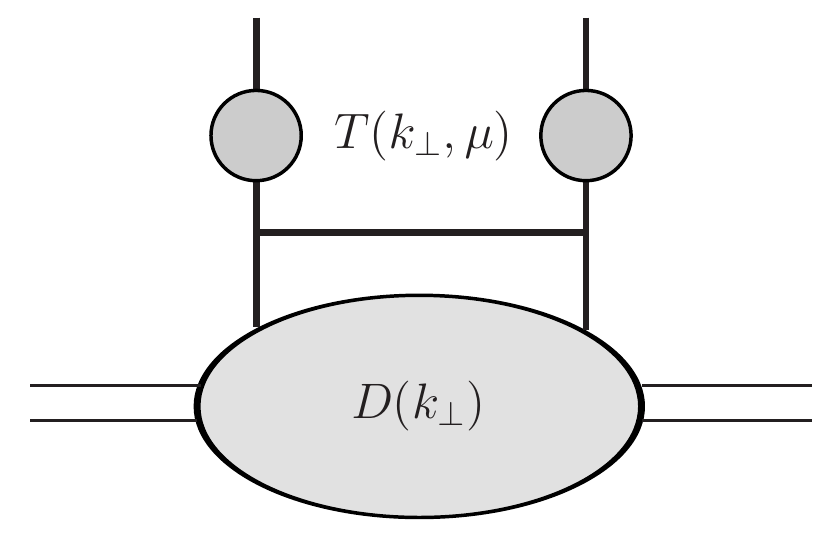}
\caption{Schematic representation of the UPDFs, Eq.~\eqref{eq:7}.  The longitudinal momentum structure is suppressed on this plot. The horizontal line denotes the real parton emission with the splitting functions $P_{ab}$
and the circular blobs on the vertical lines indicate the Sudakov form factor.}\label{fig:fig1}
\end{center}
\end{figure} 

Let us take for the factorization scale parton transverse momentum, $\mu=|{\bf k}_\perp|\equiv\kperp$ and   rewrite these equations in the following form
\be
\label{eq:2}
\frac{\partial{D_a(x,\kperp)}}{\partial \ln \kperp^2} + D_a(x,\kperp)\sum_{\aprime}\int_0^{1-\Delta} dz z\,P_{\aprime a}(z,\kperp)
= \sum_{\aprime}\int_x^{1-\Delta} \frac{dz}{z}\,
P_{a\aprime}(z,\kperp)\, D_{\aprime}\Big(\frac{x}{z},\kperp\Big)\,.
\ee
After multiplying both sides of this equation by
the Sudakov form factor\footnote{Due to relations between the LO splitting functions, one power of $z$ under the integral can be removed at the price of introducing an overall factor $1/2$ in the argument of the exponent.}\!,
\be
\label{eq:3}
T_a(Q,\kperp) =\exp\left\{
-\int_{\kperp^2}^{Q^2} \frac{d\pperp^2}{\pperp^2}\sum_{\aprime}\int_0^{1-\Delta} dz z P_{\aprime a}(z,\pperp)
\right\},
\ee
where $\kperp^2\le Q^2$, 
the l.h.s. can be written as a full derivative, and Eq.~(\ref{eq:2}) reads
\be
\label{eq:4}
\frac{\partial}{\partial \ln \kperp^2}\left[T_a(Q,\kperp) D_a(x,\kperp) \right]=
T_a(Q,\kperp)\sum_{\aprime} \int_x^{1-\Delta} \frac{dz}{z}\,
P_{a\aprime}(z,\kperp) D_{\aprime}\Big(\frac{x}{z},\kperp\Big)\,.
\ee
The Sudakov form factor is interpreted as the probability  that the parton $a$ with transverse momentum $\kperp$ will not split into a pair of partons during the evolution in $\pperp^2$ up to the scale $Q^2$. 
Integrating both sides of Eq.~(\ref{eq:4}) over $\kperp$ in the limits $Q_0\le  Q$,  where $Q_0$ is an initial scale for the DGLAP evolution, 
we find on the l.h.s.
\be
\label{eq:5}
\int_{Q_0^2}^{Q^2}\frac{d\kperp^2}{\kperp^2}\,\frac{\partial}{\partial \ln \kperp^2}\left[T_a(Q,\kperp) D_a(x,\kperp) \right]
= D_a(x,Q)-T_a(Q,Q_0) D_a(x,Q_0)\,,
\ee
since $T_a(Q,Q)=1$. Thus, Eq.~(\ref{eq:4}) takes the following form
\be
\label{eq:6}
D_a(x,Q) = T_a(Q,Q_0,) D_a(x,Q_0) +  \int_{Q_0^2}^{Q^2}\frac{d\kperp^2}{\kperp^2}\bigg\{
T_a(Q,\kperp)\sum_{\aprime}\int_x^{1-\Delta} \frac{dz}{z}\,P_{a\aprime}(z,\kperp)\, 
D_{\aprime}\Big(\frac{x}{z},\kperp\Big)\bigg\}\,.
\ee
The first term on the r.h.s. corresponds to the absence of  splitting during the evolution from $Q_0$ to $Q$ while the second one describes a sequence of partonic emissions interlaced with the probabilities for no emissions. 
This constitutes the Monte Carlo scheme for generation of parton cascades.  

The expression in the curly brackets in  Eq.~(\ref{eq:6}) defines the unintegrated parton distribution functions (UPDFs), 
\be
\label{eq:7}
f_a(x,\kperp,Q) \equiv 
T_a(Q,\kperp)\sum_{\aprime}\int_x^{1-\Delta} \frac{dz}{z}\,P_{a\aprime}(z,\kperp)\, 
D_{\aprime}\Big(\frac{x}{z},\kperp\Big)\,,
\ee
defined for each flavor $a$ (including gluon).
The  transverse momentum structure of the above equation is illustrated in Fig.~\ref{fig:fig1}.
Notice that the UPDFs can also be written as a derivative 
\be
\label{eq:8}
f_a(x,\kperp,Q)= \frac{\partial}{\partial\ln \kperp^2}   \left[T_a(Q,\kperp) D_a(x,\kperp)\right].
\ee
The UPDFs in the presented scheme are defined for the transverse momenta in the range $Q_0\le\kperp \le Q$. The region below $Q_0$ merges into the non-perturbative domain and is effectively described by the initial distribution $D_a(x,Q_0)$ in the first term on the r.h.s. of Eq.~(\ref{eq:6}).
For small values of $x$,  parton saturation effects become important in  this region of transverse momenta and  special attention 
is necessary 
in phenomenological approaches to the description of physical processes in this kinematic region, see e.g. \cite{Kutak:2011fu}. The discussion of such effects, however,  is beyond the scope of the present paper.

In order to fully fix the UPDFs, the cutoff parameter $\Delta$ in Eqs.~(\ref{eq:3}) and (\ref{eq:7}) needs to be specified. 
In Ref.~\cite{Kimber:1999xc} 
the cutoff   was set in the spirit of the DGLAP ordering of parton real emission in transverse momenta to
\be
\label{eq:9a}
\Delta =\frac{\kperp}{Q}\,.
\ee
Thus, from the upper integration limit, $x<(1-\Delta)$, 
the UPDFs are nonzero for $\kperp < Q(1-x)$.
The Sudakov form factor $T_a(Q,Q_0)$ in Eq.~\eqref{eq:6} is also regulated with the corresponding cutoff $\Delta_0=Q_0/Q$.

The prescription was further modified in Ref.~\cite{Kimber:2000bg,Kimber:2001sc} to account for the angular ordering in parton emissions in accord with the CCFM evolution scheme \cite{Ciafaloni:1987ur,Catani:1989yc,Catani:1989sg,Marchesini:1994wr},
\be
\label{eq:9b}
\Delta=\frac{\kperp}{\kperp+Q}\,.
\ee
In such a case, the nonzero values of the UDPFs are given for 
 $ \kperp< Q({1}/{x}-1)$. The upper cutoff now is bigger  than in the DGLAP scheme. This is particularly important for  the small $x$ values  which  allows for a smooth transition of transverse momenta in the CCFM scheme into the region $\kperp \gg Q$, see Ref.~\cite{Kimber:2000bg,Kimber:2001sc} for more details.

\subsection{Alternative derivation}
\label{sec:ad}

In  this subsection we shall construct an alternative derivation of the unintegrated single parton density. The aim is to prepare the ground and methods for the construction of the unintegrated double parton distributions.
The solution of the DGLAP equations (\ref{eq:1}) can be written in terms of the parton-to-parton evolution distributions $E_{ab}(x,\mu_0,\mu)$, which obey the following equation
\be
\label{eq:b12}
\frac{\partial}{\partial \ln \mu^2} E_{ab}(x,\mu,\mu_0)=\sum_{\aprime}\int_x^{1}\frac{dz}{z}\,P_{a\aprime}(z,\mu)\,
E_{\aprime b}\Big(\frac{x}{z},\mu,\mu_0\Big)
-E_{ab}(x,\mu,\mu_0)\,\sum_{\aprime}\int_0^{1}dz z P_{\aprime a}(z,\mu) \;,
\ee
with the initial condition
\be
E_{ab}(x,\mu_0,\mu_0)=\delta_{ab}\,\delta(1-x)\,.
\ee
In the above we have  regularized singularity of the splitting functions at $z=1$ by introducing a small parameter $\epsilon$, e.g. 
$P^{(0)}_{qq}(z) \sim {1}/{(1-z+\epsilon)}$
for $z\to 1$.
These distributions generate the evolution of the PDFs
\be
\label{eq:b13}
D_a(x,\mu)=\sum_b\int_x^1\frac{dz}{z}\,E_{ab}\Big(\frac{x}{z},\mu,\mu_0\Big)\, D_b(z,\mu_0)\,,
\ee
since the parton distributions obtained from this relation obey the standard DGLAP evolution equations.
This can be easily proven by using the Mellin transform
\be
\label{eq:b14}
\Atilde(n)=\int_0^{1}dx\,x^nA(x)\,.
\ee
Using the above definition,  Eq.~(\ref{eq:b12})   reads
\be
\label{eq:b15}
\frac{\partial}{\partial \ln \mu^2} \Etilde_{ab}(n,\mu,\mu_0)=\sum_{\aprime}\Ptilde_{a\aprime}(n,\mu)\,\Etilde_{\aprime b}(n,\mu,\mu_0)
-\Etilde_{ab}(n,\mu,\mu_0)\,\sum_{\aprime}\int_0^1dz z P_{\aprime a}(z,\mu)\; ,
\ee
with the initial condition $\Etilde_{ab}(n,\mu_0,\mu_0)=\delta_{ab}$, while Eq.~(\ref{eq:b13})  is given by
\be
\label{eq:b16}
\Dtilde_a(n,\mu)=\sum_b \Etilde_{ab}(n,\mu,\mu_0)\,\Dtilde_b(n,\mu_0)\,.
\ee
Multiplying both sides of Eq.~(\ref{eq:b15}) by $\Dtilde_b(n,\mu_0)$ and summing over $b$, we obtain Eq.~(\ref{eq:1}) 
in the Mellin moment space
\be
\label{eq:b17}
\frac{\partial}{\partial \ln \mu^2} \Dtilde_{a}(n,\mu)=\sum_{\aprime}\Ptilde_{a\aprime}(n,\mu)\,\Dtilde_{\aprime}(n,\mu)
-\Dtilde_{a}(n,\mu)\,\sum_{\aprime}\int_0^1dz z P_{\aprime a}(z,\mu)\,.
\ee

To find the UPDFs, we  set $\mu=\kperp$ in  Eq.~(\ref{eq:b15}) and multiply both sides by the Sudakov form factor,
\be
\label{eq:b21}
T_a(Q,\kperp)=\exp\bigg\{-\int_{\kperp^2}^{Q^2}\frac{d\pperp^{2}}{\pperp^{2}}\sum_{\aprime}\int_0^1dz z P_{\aprime a}(z,\kperp)\bigg\},
\ee
to obtain
\be
\label{eq:b22}
\frac{\partial}{\partial \ln \kperp^2} \left[T_a(Q,\kperp)\Etilde_{ab}(n,\kperp,\mu_0)\right]=
T_a(Q,\kperp)\sum_{\aprime}\Ptilde_{a\aprime}(n,\kperp)\,\Etilde_{\aprime b}(n,\kperp,\mu_0)\,.
\ee
Integrating both sides of this equation over $\kperp$ from $\mu_0\equiv Q_0$ to $Q$, we find
\be
\label{eq:b23}
\Etilde_{ab}(n,Q,Q_0) = T_a(Q,Q_0)\,\delta_{ab}+
\int_{Q_0^2}^{Q^2}  \frac{d\kperp^2}{\kperp^2}\,T_a(Q,\kperp)
\sum_{\aprime}\Ptilde_{a\aprime}(n,\kperp)\,\Etilde_{\aprime b}(n,\kperp,Q_0)\,,
\ee
and using Eq.~(\ref{eq:b16}), we obtain
\be
\label{eq:b24}
\Dtilde_a(n,Q) = T_a(Q,Q_0)\Dtilde_a(n,Q_0)+
\int_{Q_0^2}^{Q^2}  \frac{d\kperp^2}{\kperp^2}\,T_a(Q,\kperp)
\sum_{\aprime}\Ptilde_{a\aprime}(n,\kperp)\,\Dtilde_{\aprime}(n,\kperp)\,.
\ee
The expression under the integral in the above equation is the unintegrated parton distribution  in the Mellin moment space.
Transforming it  into the $x$ space, we find the following equation
\be
\label{eq:b26}
f_a(x,\kperp,Q) =T_a(Q,\kperp)\,\sum_{\aprime}\int_x^{1} \frac{dz}{z}\,P_{a\aprime}(z,\kperp)\, D_{\aprime}\Big(\frac{x}{z},\kperp\big)\,,
\ee
which is equivalent to  Eq.~(\ref{eq:7})  after switching from the $\epsilon$ regularization of the splitting functions  to the regularization with $(1-\Delta)$ in the upper integration limit, both in   the above equation and in the Sudakov form factor (\ref{eq:b21}).

\section{Double parton distributions}
\label{sec:dpdf}

We start this section from recalling the evolution equations for the  integrated double parton distribution functions,
$D_{a_1a_2}(x_2,x_2,Q_1,Q_2)$, following results of Ref.~\cite{Ceccopieri:2010kg} appended by virtual corrections\footnote{The DPDFs
also depend  on the transverse momentum vector $\rperpb$, which we set to zero. For $\rperpb=0$, the DPDFs in the lowest order approximation are probabilities to find two partons with longitudinal momentum fractions $x_{1,2}$, see \cite{Diehl:2011yj} for  more details.}. 
 The evolution of the DPDFs can be cast in the following form
\begin{align}\nonumber
\label{eq:9}
D_{a_1a_2}(x_1,x_2, &\, Q_1,Q_2) =
\sum_{\aprime,\abis}\bigg\{
\int_{x_1}^{1-x_2}\frac{dz_1}{z_1}\int_{x_2}^{1-z_1}\frac{dz_2}{z_2}\,
E_{a_1\aprime}\Big(\frac{x_1}{z_1},Q_1,Q_0\Big)
E_{a_2\abis}\Big(\frac{x_2}{z_2},Q_2,Q_0\Big)D_{\aprime \abis}(z_1,z_2,Q_0,Q_0)
\\\nonumber
\\
&+
\int^{Q_{\min}^2}_{Q_0^2}\frac{dQ^2_s}{Q^2_s}
\int_{x_1}^{1-x_2}\frac{dz_1}{z_1}
\int_{x_2}^{1-z_1}\frac{dz_2}{z_2}\,
E_{a_1\aprime}\Big(\frac{x_1}{z_1},Q_1,Q_s\Big)
E_{a_2\abis}\Big(\frac{x_2}{z_2},Q_2,Q_s\Big)D_{\aprime \abis}^{(sp)}(z_1,z_2,Q_s)\bigg\}\; ,
 \end{align}
where $Q_{\min}^2=\min\{Q_1^2,Q_2^2\}$, and the distributions $E_{ab}$ obey the evolution equation (\ref{eq:b12}).
The integration limits take into account kinematic constraints $x_1,x_2>0$ and $x_1+x_2\le 1$.  

\begin{figure}[t]
\begin{center}
\includegraphics[width=0.22\textwidth]{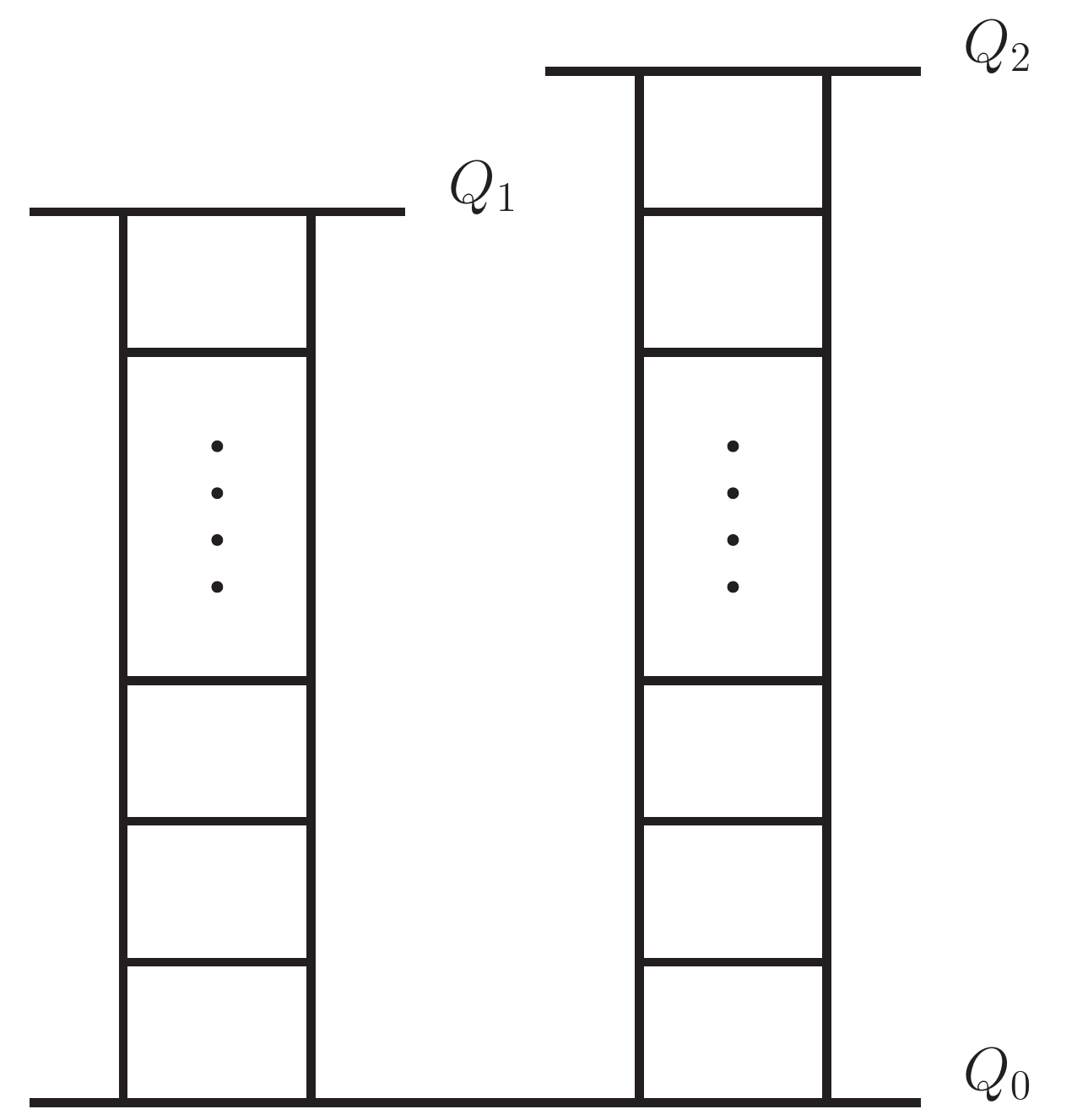}\hspace*{1.7cm}
\includegraphics[width=0.22\textwidth]{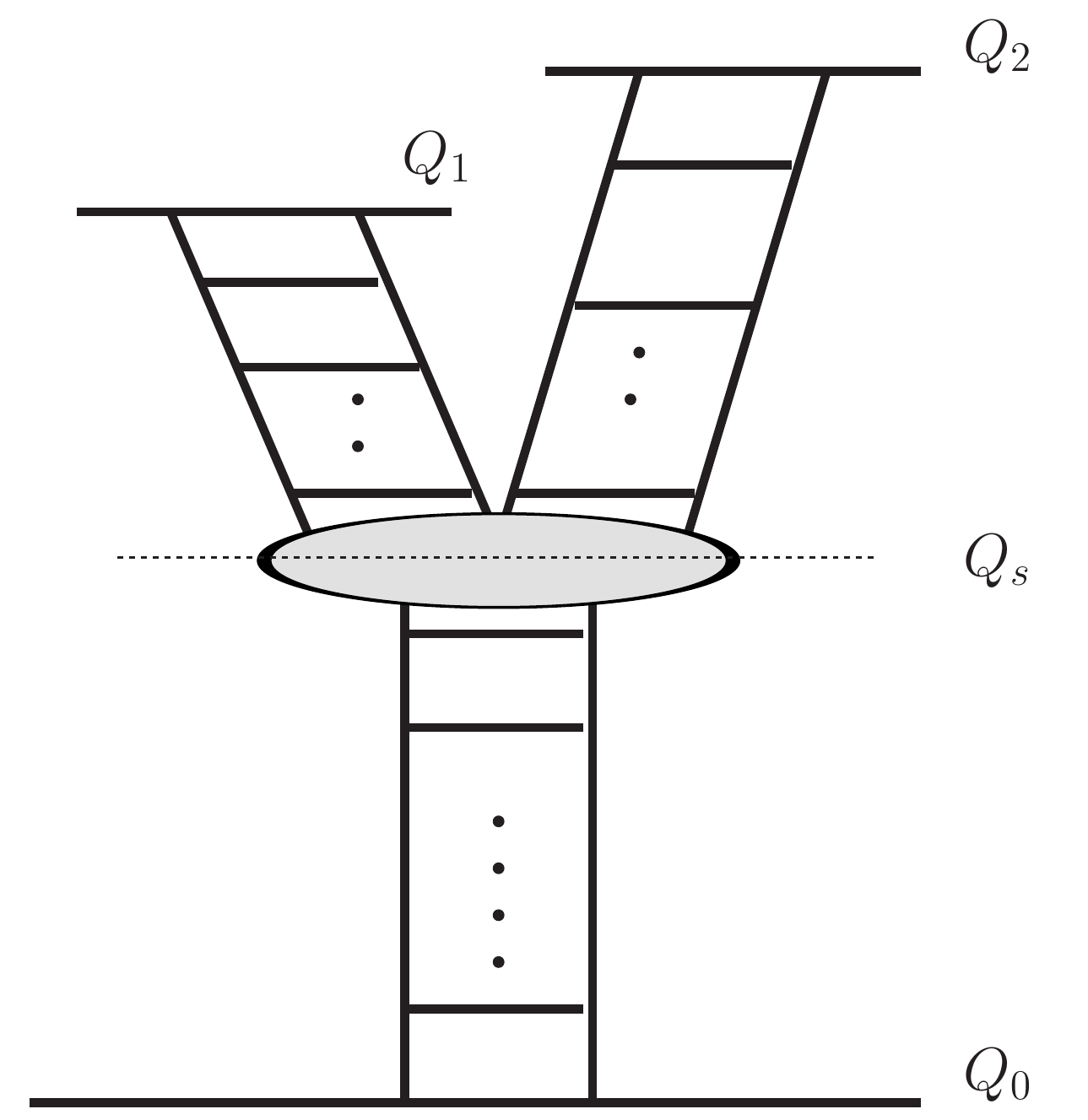}
\vspace*{0.5cm}
\caption{Schematic illustration of  the two contributions to the DPDFs (\ref{eq:9}). Left: homogeneous term; right: inhomogeneous term. It is understood that all the ladders are cut diagrams.}
\label{fig:fig1a}
\end{center}
\end{figure}

The first,  homogenous, term on the r.h.s. of Eq.~\eqref{eq:9}, is proportional to the double parton density and corresponds to the independent evolution of two partons  from the initial scale $Q_0$ to $Q_1$ and from $Q_0$ to $Q_2$. The second, non-homogeneous, term
contains the distribution
\be
\label{eq:10}
D_{\aprime \abis}^{(sp)}(x_1,x_2,Q_s) =\frac{\alpha_s(Q_s)}{2\pi}\sum_{a}\frac{D_{a}(x_1+x_2,Q_s)}{x_1+x_2}\,
P_{a\to \aprime \abis}\bigg(\frac{x_1}{x_1+x_2}\bigg)\,,
\ee
which describes the splitting of the parton $a\to \aprime\abis$. Notice the  single PDFs, $D_{a}$,
 at the splitting scale $Q_s$  along with the real emission LO splitting functions (\ref{eq:1a}),
$P_{a\to \aprime \abis}(z)= P^{(0)}_{\aprime a}(z)$ on the r.h.s. .
In the LO,  the second parton flavor $\abis$ is uniquely determined from the splitting $a\to \aprime$. The two contributions in Eq.~(\ref{eq:9}) are
schematicaly shown in Fig.~\ref{fig:fig1a}.

The presented results can also be written in the Mellin moment space introducing the double Mellin transform
\be
\label{eq:18}
\Atilde(n_1,n_2)=\int_0^1dx_1\int_0^1dx_2\,x_1^{n_1}x_2^{n_2}\,\theta(1-x_1-x_2)\,A(x_1,x_2)\,.
\ee
Then Eq.~(\ref{eq:9}) reads
\begin{align}\nonumber
\label{eq:19}
\Dtilde_{a_1a_2}(n_1,n_2,Q_1,Q_2) &= \sum_{\aprime,\abis}\bigg\{
\Etilde_{a_1\aprime}(n_1,Q_1,Q_0)\,\Etilde_{a_2\abis}(n_2,Q_2,Q_0)\,
\Dtilde_{\aprime\abis}(n_1,n_2,Q_0,Q_0) 
\\
&+ \int^{Q_{\min}^2}_{Q_0^2}\frac{dQ^2_s}{Q^2_s}\,
\Etilde_{a_1\aprime}(n_1,Q_1,Q_s)\,\Etilde_{a_2\abis}(n_2,Q_2,Q_s)\,
\Dtilde^{(sp)}_{\aprime\abis}(n_1,n_2,Q_s)\bigg\} \; ,
\end{align}
and
\be
\label{eq:19b}
\Dtilde^{(sp)}_{\aprime\abis}(n_1,n_2,Q_s)=\frac{\alpha_s(Q_s)}{2\pi}\sum_a \Dtilde_a(n_1+n_2,Q_s)
\int_0^1dzz^{n_1}(1-z)^{n_2}P_{a\to\aprime\abis}(z)\,.
\ee

In the  above formula  the two hard scales are not necessary equal. In the case of equal scales, $Q_1=Q_1\equiv Q$,
Eq.~(\ref{eq:19}) is a solution to the following differential equation
\begin{align}\nonumber
\label{eq:24}
\frac{\partial}{\partial \ln Q^2}\,\Dtilde_{a_1a_2}(n_1,n_2,Q,Q) &=
\sum_{\aprime}P_{a_1\aprime}(n_1,Q)\,\Dtilde_{\aprime a_2}(n_1,n_2,Q,Q)-\Dtilde_{a_1a_2}(n_1,n_2,Q,Q)
\sum_{\aprime}\int_0^1dzzP_{\aprime a_1}(z,Q)
\\\nonumber
\\\nonumber
&+
\sum_{\aprime} P_{a_2\aprime}(n_2,Q)\,\Dtilde_{a_1\aprime}(n_1,n_2,Q,Q)-\Dtilde_{a_1a_2}(n_1,n_2,Q,Q)\sum_{\aprime}
\int_0^1dzzP_{\aprime a_2}(z,Q)
\\\nonumber
\\
&+
\Dtilde^{(sp)}_{a_1a_2}(n_1,n_2,Q)\,,
\end{align}
see Appendix A for the proof. The above equations are the well known evolution equations  
\cite{Snigirev:2003cq,Korotkikh:2004bz,Gaunt:2011xd} in the Mellin moment space.

\section{Unintegrated double parton distributions}
\label{sec:udpdf}

In this section we shall define the unintegrated double parton distribution functions by essentially generalizing the procedure introduced in \cite{Kimber:1999xc,Kimber:2001sc} for the single PDFs, reviewed in Sec.~\ref{sec:skmr}. In what follows, we shall discuss the homogeneous 
and non-homogeneous parts separately as their treatment in the presence of the transverse momentum dependence
is rather different.

\subsection{Homogeneous part in the Mellin space}

\begin{figure}[t]
\begin{center}
\includegraphics[width=0.30\textwidth]{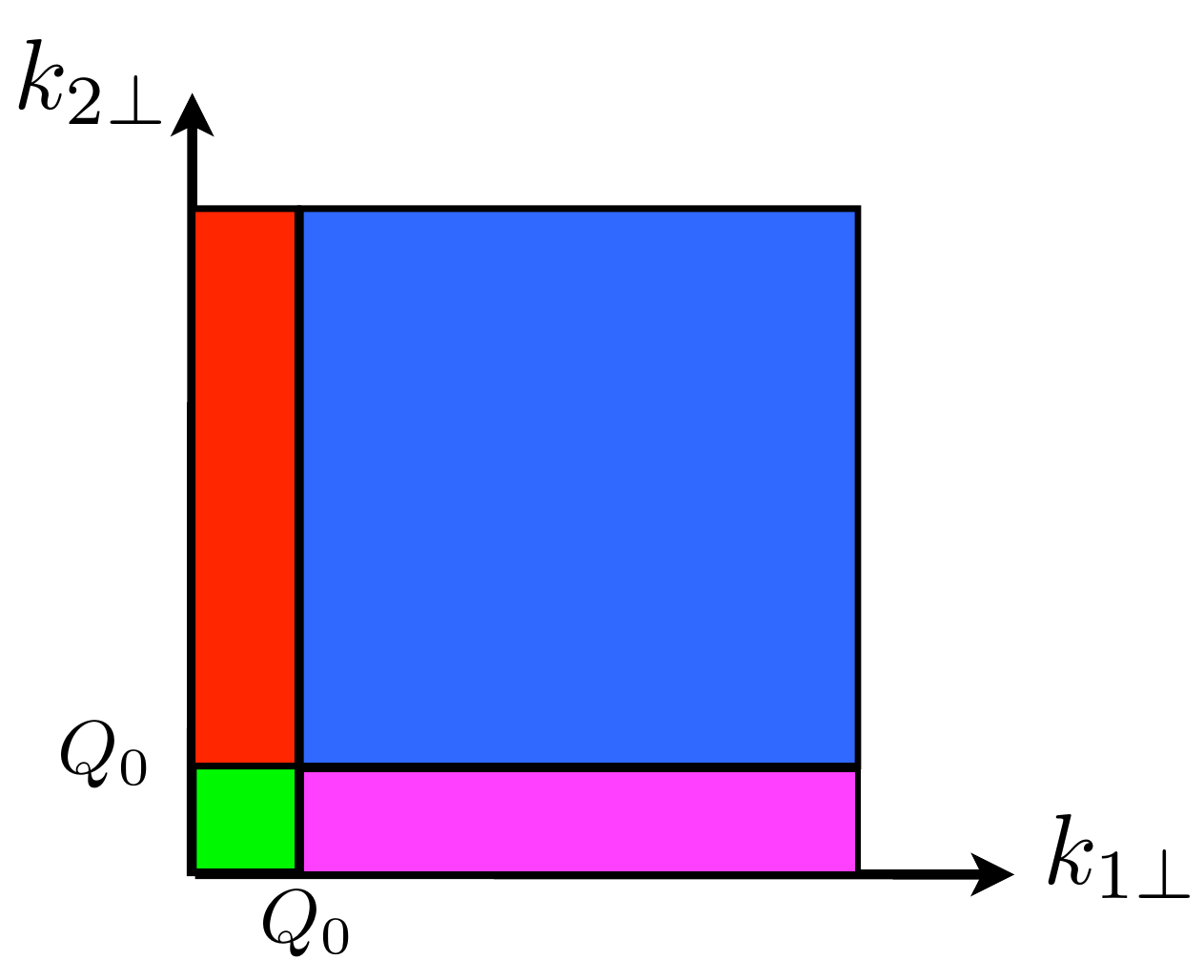}\hspace*{0.7cm}
\caption{Different regions of the transverse momenta contributing to  Eq.~\eqref{eq:ad21}. The smallest region with the transverse momenta below $Q_0$ corresponds to the first term in Eq.~\eqref{eq:ad21}. In this region the transverse momenta have been integrated over in the corresponding expression.  The regions for which either $\kperpone\le Q_0$ or $\kperptwo\le Q_0$ correspond to the two subsequent terms in  Eq.~\eqref{eq:ad21}. In this case the smaller transverse momentum has been integrated over. The largest region, with $\kperpone,\kperptwo>Q_0$, corresponds to the last term in Eq.~\eqref{eq:ad21}.}
\label{fig:fig2}
\end{center}
\end{figure}

Let us first concentrate on the homogeneous part of Eq.~(\ref{eq:19}),
\be
\label{eq:d20}
\Dtilde^{(h)}_{a_1a_2}(n_1,n_2,Q_1,Q_2) \eq\sum_{\aprime,\abis}
\Etilde_{a_1\aprime}(n_1,Q_1,Q_0)\,\Etilde_{a_2\abis}(n_2,Q_2,Q_0)\,
\Dtilde_{\aprime\abis}(n_1,n_2,Q_0,Q_0) \,.
\ee
Using Eq.~(\ref{eq:b23}) with the regularized splitting functions in the above equation, 
\begin{align}\nonumber
\label{eq:d20a}
\Dtilde^{(h)}_{a_1a_2}(n_1,n_2, Q_1,Q_2) &= \sum_{\aprime,\abis}
\bigg\{
T_{a_1}(Q_1,Q_0)\,\delta_{{a_1}\aprime}+
\int_{Q_0^2}^{Q_1^2}  \frac{d\kperpone^2}{\kperpone^2}\,T_{a_1}(Q_1,\kperpone)
\sum_{b}\Ptilde_{{a_1}b}(n_1,\kperpone)\,\Etilde_{b \aprime}(n_1,\kperpone,Q_0)
\bigg\}
\\\nonumber
\\\nonumber
&\times
\bigg\{
T_{a_2}(Q_2,Q_0)\,\delta_{a_2\abis}+
\int_{Q_0^2}^{Q_2^2}  \frac{d\kperptwo^2}{\kperptwo^2}\,T_{a_2}(Q_2,\kperptwo)
\sum_{b}\Ptilde_{{a_2}b}(n_2,\kperptwo)\,\Etilde_{b \abis}(n_2,\kperptwo,Q_0)
\bigg\}
\\\nonumber
\\
&\times
\Dtilde_{\aprime\abis}(n_1,n_2,Q_0,Q_0)\,.
\end{align}
and multiplying term by term, we obtain
\begin{align}\nonumber
\label{eq:ad21}
\Dtilde^{(h)}_{a_1a_2}(&\,n_1,n_2,Q_1,Q_2) =T_{a_1}(Q_1,Q_0)\,T_{a_2}(Q_2,Q_0)\,\Dtilde_{a_1a_2}(n_1,n_2,Q_0,Q_0)
\\\nonumber
\\\nonumber
&+ \int_{Q_0^2}^{Q_2^2}  \frac{d\kperptwo^2}{\kperptwo^2}
\bigg\{T_{a_1}(Q_1,Q_0)\,T_{a_2}(Q_2,\kperptwo)\sum_{b}\Ptilde_{a_2b}(n_2,\kperptwo)\Big[\sum_{\abis}\Etilde_{b \abis}(n_2,\kperptwo,Q_0)
\Dtilde_{a_1\abis}(n_1,n_2,Q_0,Q_0)\Big]\bigg\}
\\\nonumber
\\\nonumber
&+ \int_{Q_0^2}^{Q_1^2}  \frac{d\kperpone^2}{\kperpone^2}
\bigg\{T_{a_1}(Q_1,\kperpone)\,T_{a_2}(Q_2,Q_0)\sum_{b}\Ptilde_{a_1b}(n_1,\kperpone)\Big[\sum_{\aprime}\Etilde_{b \aprime}(n_1,\kperpone,Q_0)
\Dtilde_{\aprime a_2}(n_1,n_2,Q_0,Q_0)\Big]\bigg\}
\\\nonumber
\\\nonumber
&+
\int_{Q_0^2}^{Q_1^2}  \frac{d\kperpone^2}{\kperpone^2}
\int_{Q_0^2}^{Q_2^2}  \frac{d\kperptwo^2}{\kperptwo^2}
\bigg\{T_{a_1}(Q_1,\kperpone)\,T_{a_2}(Q_2,\kperptwo)
\sum_{b,c}\Ptilde_{a_1b}(n_1,\kperpone)\,\Ptilde_{a_2c}(n_2,\kperptwo)
\\
&~~~~~~~~~~~~~~~~~~~~~~~~~~~~~\times
\Big[\sum_{\aprime,\abis}\Etilde_{b \aprime}(n_1,\kperpone,Q_0)
\Etilde_{c \abis}(n_2,\kperptwo,Q_0)
\Dtilde_{\aprime\abis}(n_1,n_2,Q_0,Q_0)\Big]\bigg\}\,.
\end{align}
The four terms in the above equation are defined over four distinct regions of the transverse momenta $\kperpone,\kperptwo$ which are schematically illustrated in Fig.~\ref{fig:fig2}.

The first term in the sum on the r.h.s. of  Eq.~\eqref{eq:ad21} corresponds to the evolution of the two partons from the initial scale
$Q_0$ to the hard scales, $Q_1$ and $Q_2$, without real parton emissions. This term  is defined in the region of the lowest transverse momenta, $\kperpone,\kperptwo \le Q_0$, and  does not contain any transverse momentum dependence since it has been integrated out in this non-perturbative region.

The expressions in the curly brackets under the integrals in Eq.~({\ref{eq:ad21}})
are the {\it unintegrated double parton distribution functions} (UDPDFs), $\ftilde_{a_1a_2}^{(h)}$, defined in the three  remaining regions of transverse momenta.
Notice that, the expressions in the square brackets are the homogeneous DPDFs (\ref{eq:d20}), taken at appropriate scales. 
For example, in the square brackets under the first integral, we have
\be
\Dtilde^{(h)}_{a_1b}(n_1,n_2,Q_0,\kperptwo)=\sum_{\aprime,\abis}\Etilde_{a_1 \aprime}(n_1,Q_0,Q_0)\,
\Etilde_{b \abis}(n_2,\kperptwo,Q_0)\,\Dtilde_{\aprime\abis}(n_1,n_2,Q_0,Q_0)\,,
\ee
since $\Etilde_{a_1 \aprime}(n_1,Q_0,Q_0)=\delta_{a_1\aprime}$.
Thus, for $\kperpone\le Q_0$ and $\kperptwo > Q_0$, we find from the first integral
\be
\label{eq:d22a}
\ftilde_{a_1a_2}^{(h)}(n_1,n_2, \kperptwo, Q_1,Q_2)= T_{a_1}(Q_1,Q_0)\,T_{a_2}(Q_2,\kperptwo)\sum_{b}\,
\Ptilde_{a_2b}(n_2,\kperptwo)\,\Dtilde^{(h)}_{a_1b}(n_1,n_2,Q_0,\kperptwo)\,.
\ee
The dependence of the transverse momentum $\kperpone$ is integrated over up to $Q_0$ in such a case and $\kperpone$ is not present
among the arguments of the defined function. The effect of such an integration is hidden in the integrated DPDFs on the r.h.s.
taken at the scale $Q_0$ for the first parton.

Similarly, for   $\kperpone > Q_0$ and $\kperptwo\le Q_0$,  we have from the the second integral
\be
\label{eq:d22b}
\ftilde_{a_1a_2}^{(h)}(n_1,n_2,\kperpone, Q_1,Q_2)=T_{a_1}(Q_1,\kperpone)\,T_{a_2}(Q_2,Q_0)
\sum_{b}\Ptilde_{a_1b}(n_1,\kperpone)\,\Dtilde^{(h)}_{b a_2}(n_1,n_2,\kperpone,Q_0)\;.
\ee
Now the momentum $\kperptwo$ is integrated up to the scale $Q_0$ and only $\kperpone$ dependence is present.

\begin{figure}[b]
\begin{center}
\includegraphics[width=0.32\textwidth]{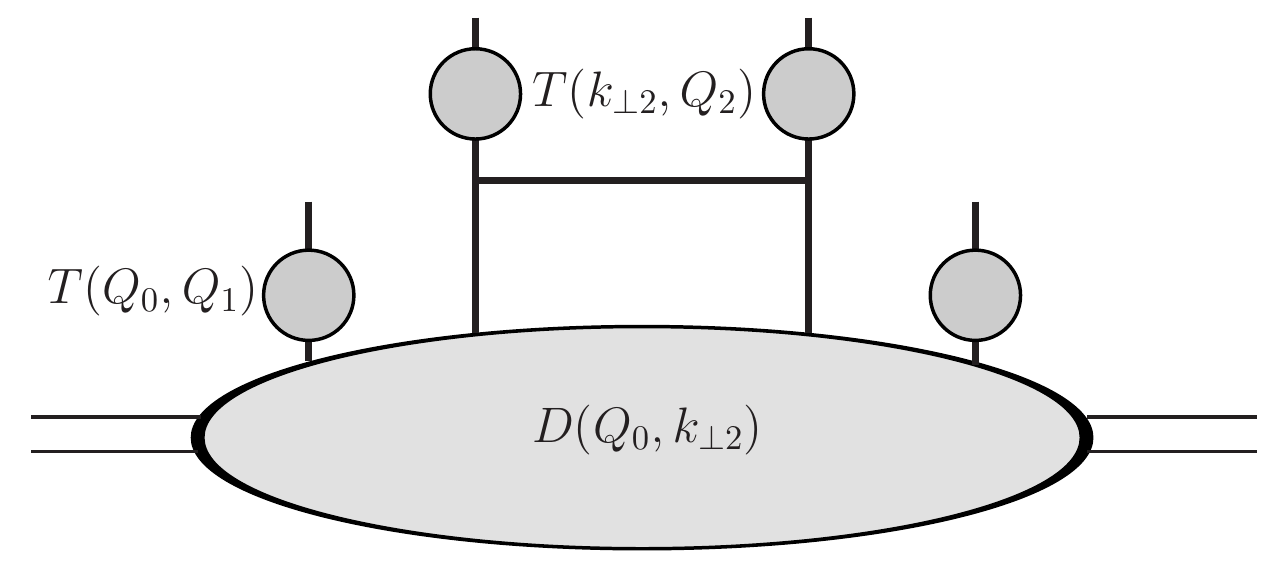}\hspace*{0.3cm}
\includegraphics[width=0.32\textwidth]{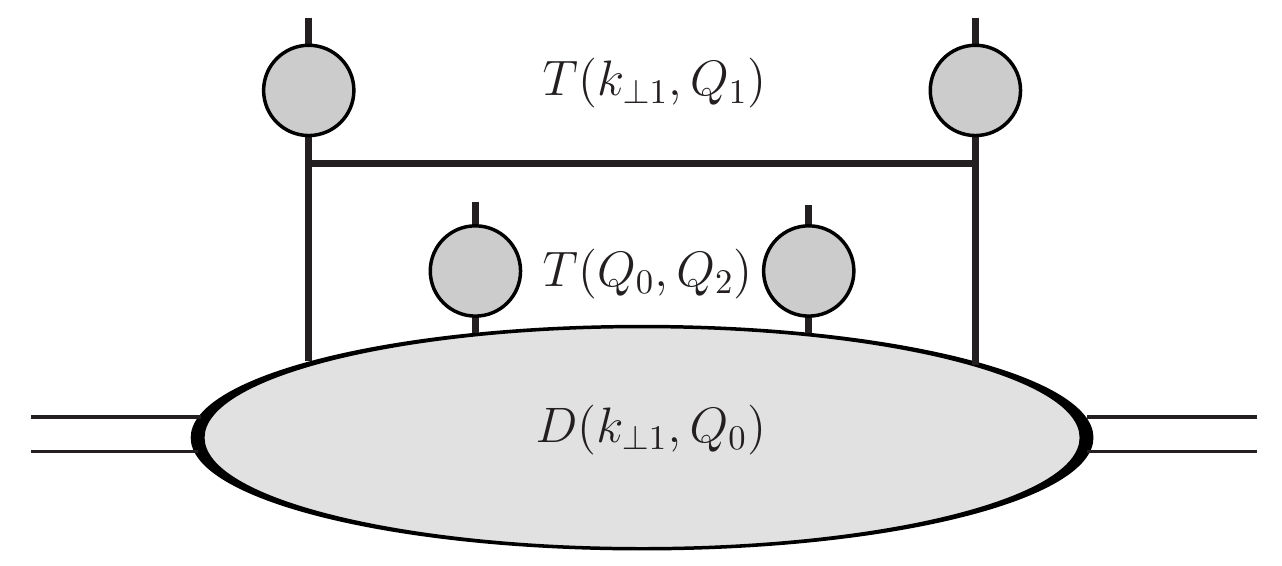}\hspace*{0.3cm}
\includegraphics[width=0.32\textwidth]{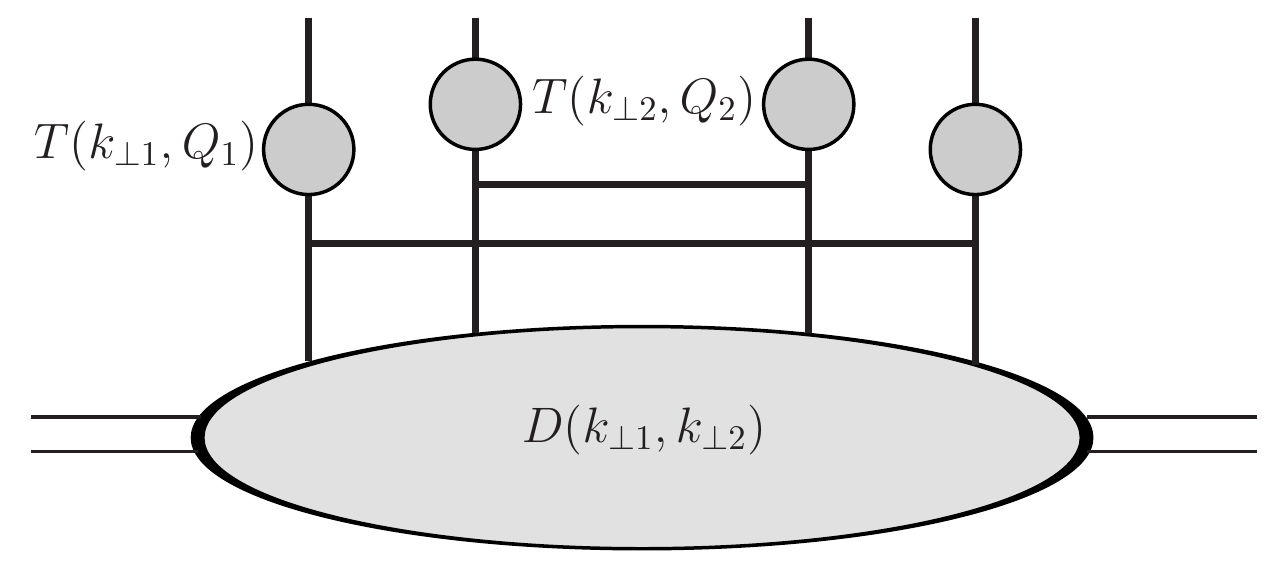}
\caption{Schematic representation of the UDPDFs 
for the  three unintegrated DPDFs given by Eqs.~(\ref{eq:d22a})-(\ref{eq:d22c}), from left to right respectively.
The  longitudinal  momentum structure is suppressed on these plots. The horizontal lines correspond to
the real parton emission with the splitting functions and the circular blobs on the vertical lines indicate the Sudakov form factors.
}\label{fig:fig3}
\end{center}
\end{figure}

Finally, for $\kperpone,\kperptwo > Q_0$  the third integral gives the dependence on both transverse momenta,
\begin{align}\nonumber
\label{eq:d22c}
\ftilde_{a_1a_2}^{(h)}(n_1,n_2,\kperpone,\kperptwo, Q_1,Q_2)& =
T_{a_1}(Q_1,\kperpone)\,T_{a_2}(Q_2,\kperptwo)\,
\\\nonumber
\\
&\times
\sum_{b,c}\Ptilde_{a_1b}(n_1,\kperpone)\,\Ptilde_{a_2c}(n_2,\kperptwo)\,
\Dtilde^{(h)}_{bc}(n_1,n_2,\kperpone,\kperptwo)\,.
\end{align}
The three unintegrated DPDFs are schematically represented in  Fig.~\ref{fig:fig3}.

In principle, all the regions of the transverse momenta need to be included for any  configuration of the external hard scales $Q_1$ and $Q_2$. It is clear though, that some regions will be subdominant depending on the scales, due to the suppression originating from the Sudakov formfactors. For example,
the first term in Eq.~\eqref{eq:ad21} is going to be very small whenever any of the scales is much larger than $Q_0$.

\subsection{Homogeneous part in  the $x$-space}

The corresponding expressions in the $x$-space can be easily found. For example, for Eq.~(\ref{eq:d22c}) we obtain
\begin{align}\nonumber
\label{eq:d23}
f_{a_1a_2}^{(h)}(x_1,x_2,&\,\kperpone,\kperptwo, Q_1,Q_2) =
T_{a_1}(Q_1,\kperpone)\,T_{a_2}(Q_2,\kperptwo)
\\\nonumber
\\
&\times\,
\sum_{b,c}\int_{x_1}^{1-x_2}\frac{dz_1}{z_1}\int_{x_2}^{1-z_1}\frac{dz_2}{z_2}
P_{a_1b}\Big(\frac{x_1}{z_1},\kperpone\Big)
P_{a_2c}\Big(\frac{x_2}{z_2},\kperptwo\Big)D^{(h)}_{bc}(z_1,z_2,\kperpone,\kperptwo)\,.
\end{align}
 Similarly to the case of the single parton distributions, the integrals over $z_{1,2}$ need to be regularized since the splitting functions can be singular for $z_1=x_1$ and $z_2=x_2$. 
So far in order to be able to manipulate these equations we have implicitly assumed the regularization through the modification of the splitting function by the parameter $\epsilon$, as indicated earlier. Following the original scheme presented in Sec.~\ref{sec:ad},  we now introduce the regularization through the cutoff in the limits of the integrals over the longitudinal momenta. 
After changing the integration variables, $z_1\to x_1/z_1$
and $z_2\to x_2/z_2$, the singularities occur for $z_{1,2}=1$, thus we change the upper integration limits form $1$ to  $1\to (1-\Delta_{1,2})$ to finally find  
\begin{align}\nonumber
\label{eq:d23a}
f_{a_1a_2}^{(h)}(x_1,x_2,&\,\kperpone,\kperptwo, Q_1,Q_2) =
T_{a_1}(Q_1,\kperpone)\,T_{a_2}(Q_2,\kperptwo)
\\\nonumber
\\
&\times\,
\sum_{b,c}
\int_{\frac{x_1}{1-x_2}}^{1-\Delta_1}\frac{dz_1}{z_1}
\int_{\frac{x_2}{1-x_1/z_1}}^{1-\Delta_2}\frac{dz_2}{z_2}
P_{a_1b}(z_1,\kperpone)\,
P_{a_2c}(z_2,\kperptwo)\,
D^{(h)}_{bc}\Big(\frac{x_1}{z_1},\frac{x_2}{z_2},\kperpone,\kperptwo\Big)\,.
\end{align}
The same regularization is necessary for the Sudakov form factors, $T_{a_{1}}$ and $ T_{a_2}$, with $\Delta_1$ and $\Delta_2$ respectively,
see Eq.~(\ref{eq:7}).
Following the presentation of the single UPDFs,  we choose 
\be
\label{eq:d24}
\Delta_i=\frac{k_{i\perp}}{Q_i}\,.
\ee

Applying the same procedure to  the rest of the UDPDFs, we find  for the distribution (\ref{eq:d22a}) 
\begin{align}\nonumber
\label{eq:d23b}
f_{a_1a_2}^{(h)}(x_1,x_2,\kperptwo, Q_1,Q_2)
&= 
 T_{a_1}(Q_1,Q_0)\,T_{a_2}(Q_2,\kperptwo)
\\\nonumber
\\
&\times  \sum_{b}\int_{\frac{x_2}{1-x_1}}^{1-\Delta_2}\frac{dz_2}{z_2}\,
P_{a_2b}(z_2,\kperptwo)\,
D^{(h)}_{a_1b}\Big(x_1,\frac{x_2}{z_2},Q_0,\kperptwo\Big).~
\end{align} 
Similarly,  we have for the distribution (\ref{eq:d22b})
\begin{align}\nonumber
\label{eq:d23c}
f_{a_1a_2}^{(h)}(x_1,x_2,\kperpone, Q_1,Q_2)
&=\,
T_{a_1}(Q_1,\kperpone)\,T_{a_2}(Q_2,Q_0)
\\\nonumber
\\
&\times
\sum_{b}
\int_{\frac{x_1}{1-x_2}}^{1-\Delta_1}\frac{dz_1}{z_1}
P_{a_1b}(z_1,\kperpone)\,D^{(h)}_{b a_2}\Big(\frac{x_1}{z_1},x_2,\kperpone, Q_0\Big)\,.~
\end{align}
Remember that the lack of the transverse momenta, $\kperpone$ or $\kperptwo$, among the arguments in the above formulas means that  they were integrated over up to the scale $Q_0$.

Eqs.~(\ref{eq:d23a})-(\ref{eq:d23c}) constitute  the main results of our analysis in the $x$-space.  
 They define the homogeneous part of the unintegrated double parton distribution functions 
in  three distinct domains of the hard scales, shown schematically in Fig.~\ref{fig:fig2}.
Similarly to the single PDF case, the transverse momentum dependence is generated by the last step in the evolution where  the distributions become dependent on the transverse momentum and the hard scales.
The three  unintegrated distributions were also discussed in \cite{Diehl:2011yj} but only for real emission. Our
results follow from a systematic derivation  with virtual corrections included. 
Notice that in the convention adopted in this paper the defined  UDPDFs are dimensionless quantities like the integrated DPDFs. 

In the presented analysis  we neglected the spin and color dependence of the double parton distributions and considered only the color singlet, spin averaged sector. For more information on this aspect see \cite{Diehl:2011yj,Diehl:2014vaa} an reference therein. We also do not consider here the  dependence on the  momentum transfer, setting it to zero.

\subsection{Correlations imposed by cutoffs}

Let us analyze whether the double integration over longitudinal momentum fractions $z_1$ and $z_2$ in Eq.~(\ref{eq:d23}) imposes
any restrictions on the choice of the cutoffs $\Delta_{1,2}>0$. The integration over $z_1$ gives meaningful  result if
\be
\label{eq:f1}
1-\Delta_1>\frac{x_1}{1-x_2}~~~~~~\Longrightarrow~~~~~~0<\Delta_1<\frac{1-x_1-x_2}{1-x_2}.
\ee
Notice that nonzero values of $\Delta_1$ exist for any value of parton momentum fractions since 
$(1-x_1-x_2)/(1-x_2)>0$. On the other hand, the limits of the integration over $z_2$ should fulfill
\be
\label{eq:f2}
1-\Delta_2>\frac{x_2}{1-x_1/z_1}~~~~~~\Longrightarrow~~~~~~0<\Delta_2<1-\frac{x_2}{1-x_1/z_1},
\ee
and the nonzero value of $\Delta_2$ is allowed  if
\be
\label{eq:f3}
1-\frac{x_2}{1-x_1/z_1}>0~~~~~~\Longrightarrow~~~~~~z_1>\frac{x_1}{1-x_2} \; .
\ee
The last condition is always true, which implies that for any fixed values of $x_{1,2}$, the nonzero range of the cutoff values  is  possible. This means that
with the choice  (\ref{eq:d24}), the transverse momenta of partons are bounded between zero and some maximal values
which depend on $x_{1,2}$ and $Q_{1,2}$.

With given cutoffs $\Delta_{1,2}$, the nonzero  UDPDFs are defined in a  region of $x_{1,2}$ values which are smaller
than that defined by the usual conditions, $x_{1,2}>0$ and $x_1+x_2\le 1$. In particular, 
 Eq.~(\ref{eq:f2}) constrains the lower limit of the $z_1$ integration, 
\be
\label{eq:f4}
\frac{x_1}{1-x_2/(1-\Delta_2)} \,\le\, z_1 \,\le\, 1-\Delta_1,
\ee
which  leads to the following condition
\be
\label{eq:f5}
1-\frac{x_1}{1-\Delta_1}-\frac{x_2}{1-\Delta_2}\ge 0.
\ee
The region defined by the above condition is indeed smaller than the standard one, $(1-x_1-x_2)\ge 0$.
In view of these results, with the transverse momentum dependent cutoffs (\ref{eq:d24}), the variables $(x_1,x_2,\kperpone,\kperptwo, Q_1, Q_2)$ are strongly  correlated  in the UDPDFs.

\section{Non-homogeneous part of UDPDFs}
\label{sec:nonhom}

We shall now turn to the discussion of the inhomogeneous term in the parton evolution. As we shall see,  the inclusion of the transverse momentum dependence for this contribution   is much more complicated than for the homogeneous part.
This is because, there is another source of the transverse momentum dependence in the double parton distributions. The  parent parton can perturbatively split into two daughter partons with  transverse momenta $\kperpone, \kperptwo \ge Q_0$. This mechanism is a source of parton correlations which is purely perturbative.

\subsection{Transverse momenta from evolution of two partons}

Let us consider the non-homogeneous part of Eq.~(\ref{eq:19}) which describes the splitting contribution,
\be
\label{eq:c1}
\Dtilde_{a_1a_2}^{(nh)}(n_1,n_2,Q_1,Q_2) =\int^{Q_{\min}^2}_{Q_0^2}\frac{dQ_s^2}{Q_s^2}\,\sum_{\aprime,\abis}\Etilde_{a_1\aprime}(n_1,Q_1,Q_s)\,\Etilde_{a_2\abis}(n_2,Q_2,Q_s)\,
\Dtilde^{(sp)}_{\aprime\abis}(n_1,n_2,Q_s)\,,
\ee
where $Q^2_{\min}={\min}\{Q_1^2,Q_2^2\}$.
We see that there are two potential sources of the transverse momentum dependence in the this formula, from the splitting vertex itself,  and  from the evolution above the splitting vertex. In this section, we will discuss the latter possibility
by applying the procedure developed for the homogeneous part of the UDPDFs. In that way we shall explicitly see the limits of the applicability of this formulation.

Due to the integration over $Q_s^2$, the splitting  contribution (\ref{eq:c1}) is sizable only in the case $Q_{1,2}^2\gg Q_0^2$, the condition we consider from now on. Substituting  Eq.~(\ref{eq:b23}) in Eq.~(\ref{eq:c1}) and multiplying the obtained expressions term by term,  we find the formula similar to that for the homogeneous part (\ref{eq:ad21}), 
\begin{align}
\nonumber
\label{eq:c3}
\Dtilde_{a_1a_2}^{(nh)}(n_1,&\,n_2,Q_1,Q_2)=
\int_{Q_0^2}^{Q_{\min}^2}\frac{dQ_s^2}{Q_s^2}
\bigg[
T_{a_1}(Q_1,Q_s)\,T_{a_2}(Q_2,Q_s)\,\Dtilde^{(sp)}_{a_1a_2}(n_1,n_2,Q_s)
\\\nonumber
\\\nonumber
&+
\int_{Q_s^2}^{Q_2^2}  \frac{d\kperptwo^2}{\kperptwo^2}
\bigg\{T_{a_1}(Q_1,Q_s)\,T_{a_2}(Q_2,\kperptwo)
\sum_{b}
\Ptilde_{a_2b}(n_2,\kperptwo)\sum_{\abis}
\Etilde_{b\abis}(n_2,\kperptwo,Q_s)\,
\Dtilde^{(sp)}_{a_1\abis}(n_1,n_2,Q_s)\bigg\}
\\\nonumber
\\\nonumber
&+
\int_{Q_s^2}^{Q_1^2}  \frac{d\kperpone^2}{\kperpone^2}\bigg\{
T_{a_2}(Q_2,Q_s)\,T_{a_1}(Q_1,\kperpone)
\sum_{b}
\Ptilde_{a_1b}(n_1,\kperpone)\sum_{\aprime}
\Etilde_{b\aprime}(n_1,\kperpone,Q_s)\,
\Dtilde^{(sp)}_{\aprime a_2}(n_1,n_2,Q_s)\bigg\}
\\\nonumber
\\\nonumber
&+
\int_{Q_s^2}^{Q_1^2}  \frac{d\kperpone^2}{\kperpone^2}
\int_{Q_s^2}^{Q_2^2}  \frac{d\kperptwo^2}{\kperptwo^2}\bigg\{
T_{a_1}(Q_1,\kperpone)\,T_{a_2}(Q_2,\kperptwo) \sum_{b,c}\Ptilde_{a_1b}(n_1,\kperpone)\,\Ptilde_{a_2c}(n_2,\kperptwo)
\\
&~~~~~~~~~~~~~~~~~~~~~~~~~~~~~\times\sum_{\aprime,\abis}
\Etilde_{b\aprime}(n_1,\kperpone,Q_s)\,
\Etilde_{c\abis}(n_2,\kperptwo,Q_s)\,
\Dtilde^{(sp)}_{\aprime\abis}(n_1,n_2,Q_s)\bigg\}\bigg]\,.
\end{align}
Notice that  the transverse
momenta  are confined to the perturbative region only, $\kperpone,\kperptwo\ge Q_0$, see the blue rectangle in Fig.~\ref{fig:fig2}. 
Introducing the  relation 
\be
\label{eq:c4}
\int_{Q_s^2}^{Q_{i}^2} \frac{dk_{i\perp }^2}{k_{i\perp }^2} =\int_{Q_0^2}^{Q_i^2} \frac{dk_{i\perp}^2}{k_{i\perp}^2}\,
\theta(k_{i\perp}^2-Q_s^2)\,,~~~~~~~~i=1,2\,,
\ee
one can change the order of the integrations over $Q_s^2$ and transverse momenta $k_{i\perp}^2$ in Eq.~(\ref{eq:c3}).
Thus, from the  third integral in the square brackets we  find the following non-homogenenous part of the UDPDFs,
\begin{align}
\nonumber
\label{eq:c5}
\ftilde_{a_1a_2}^{(nh)} (n_1,&\,n_2,\kperpone,\kperptwo,Q_1,Q_2)
=T_{a_1}(Q_1,\kperpone)\,T_{a_2}(Q_2,\kperptwo) \sum_{b,c}
\Ptilde_{a_1b}(n_1,\kperpone)\,\Ptilde_{a_2c}(n_2,\kperptwo)
\\
&\times
\int_{Q_0^2}^{Q_{\min}^2}\frac{dQ_s^2}{Q_s^2}\,
\theta(k_{1\perp}^2-Q_s^2)\,\theta(k_{2\perp}^2-Q_s^2)\,
\tilde{\cal D}^{(sp)}_{bc}(n_1,n_2,\kperpone,\kperptwo,Q_s)\,,
\end{align}
where  we defined new distributions
\be
\label{eq:c5a}
\tilde{\cal D}^{(sp)}_{bc}(n_1,n_2,\kperpone,\kperptwo,Q_s)
=\sum_{\aprime,\abis}\Etilde_{b\aprime}(n_1,\kperpone,Q_s)\, \Etilde_{c\abis}(n_2,\kperptwo,Q_s)\,
\Dtilde^{(sp)}_{\aprime\abis}(n_1,n_2,Q_s)\,.
\ee
The new distributions have the same structure as the homogeneous distributions $\Dtilde^{(h)}_{bc}$ in Eq.~(\ref{eq:d22c}), corresponding to two independent DGLAP evolutions  from the scale $Q_s$  (where the collinear splitting of a single parent parton to two daughter partons occurs) up to the scales given by the transverse momenta.  The initial conditions for such  evolutions are given by the known distributions  
$\Dtilde^{(sp)}_{\aprime\abis}(n_1,n_2,Q_s)$, defined in Eq.~(\ref{eq:19b}).

The regularized expression for the distribution (\ref{eq:c5})
in the $x$-space can be found in the same way as for the homogeneous part,
\begin{align}
\nonumber
\label{eq:c6}
f^{(nh)}_{a_1a_2}(x_1,x_2,\kperpone,\kperptwo,Q_1,Q_2) &=
T_{a_1}(Q_1,\kperpone)\,T_{a_2}(Q_2,\kperptwo)
\\\nonumber
\\\nonumber
&\times\,
\int_{\frac{x_1}{1-x_2}}^{1-\Delta_1}\frac{dz_1}{z_1}
\int_{\frac{x_2}{1-x_1/z_1}}^{1-\Delta_2}\frac{dz_2}{z_2}\,\sum_{b,c}
P_{a_1b}(z_1,\kperpone)
P_{a_2c}(z_2,\kperptwo)
\\\nonumber
\\
&\times
\int_{Q_0^2}^{Q_{\min}^2}\frac{dQ_s^2}{Q_s^2}\,
\theta(k_{1\perp}^2-Q_s^2)\,\theta(k_{2\perp}^2-Q_s^2)\,
{\cal D}^{(sp)}_{bc}\Big(\frac{x_1}{z_1},\frac{x_2}{z_2},\kperpone,\kperptwo,Q_s\Big)\,,
\end{align}
where  $\Delta_{1,2}$ are given by Eq.~(\ref{eq:d24}), and ${\cal D}^{(sp)}_{bc}$
are the distributions (\ref{eq:c5a}) transformed back to the $x$-space.

The  three remaining terms in Eq.~(\ref{eq:c3}) correspond to the situation in which one or two partons from the splitting do not evolve. For example, if the first parton does not evolve, we find the following expression,
\begin{align}
\nonumber
\label{eq:c5b}
\ftilde_{a_1a_2}^{(nh)} (n_1,n_2,\kperptwo,Q_1,Q_2)  &=
T_{a_2}(Q_2,\kperptwo)
\sum_{b}
\Ptilde_{a_2b}(n_2,\kperptwo)
\\
&\times
\int_{Q_0^2}^{Q_{\min}^2}\frac{dQ_s^2}{Q_s^2}\,
\theta(k_{2\perp}^2-Q_s^2)\,T_{a_1}(Q_1,Q_s)\,\tilde{\cal D}^{(sp)}_{a_1b}(n_1,n_2,Q_s,\kperptwo,Q_s)
\end{align}
where the new distributions on the r.h.s. now read,
\be
\label{eq:c5bb}
\tilde{\cal D}^{(sp)}_{a_1b}(n_1,n_2,Q_s,\kperptwo,Q_s)=
\sum_{\aprime,\abis}\Etilde_{a_1\aprime}(n_1,Q_s,Q_s)\, \Etilde_{b\abis}(n_2,\kperptwo,Q_s)\,
\Dtilde^{(sp)}_{\aprime\abis}(n_1,n_2,Q_s)\,,
\ee
since $\Etilde_{a_1\aprime}(n_1,Q_s,Q_s)=\delta_{a_1\aprime}$. 
Comparing  Eq.~(\ref{eq:c5b}) to its homogenenous
counterpart (\ref{eq:d22a}), we see that
in both expressions the  transverse momentum $\kperpone$ is not present. In Eq.~(\ref{eq:d22a}), the momentum $\kperpone$ is integrated out  in the non-perturbative domain, $\kperpone\le Q_0,$, thus,  we may set the first parton on shell ($\kperpone=0$)  in 
the $\kperp$-factorized cross sections with  off-shell matrix elements.
In the case of the distribution (\ref{eq:c5b}), however, the transverse momentum $\kperpone$ is integrated out in the perturbative region, $Q_0\le \kperpone\le Q_s$. Therefore, such a procedure is no longer justified and the distributions (\ref{eq:c5b}) cannot be used  in the $\kperp$-factorized cross sections. The same conclusion is valid when the second parton or both partons do not evolve.
In summary,  only the formula (\ref{eq:c6}) in the $x$-space is acceptable for the UDPDFs in the non-homogenous case.

\subsection{Parton splitting from light-front perturbation theory}

In order to address the issues of the transverse momentum dependence coming directly from the 
perturbative splitting of a single parent parton into two daughter partons,  
we shall utilize the methods of the light-front perturbation theory. 

Let us first start with the rederivation of the splitting term for the integrated parton densities using this framework. The definition of the integrated parton density using the light-front wave functions 
can be cast in the following form, (see  \cite{Kovchegov:2012mbw})
\begin{multline}
D_a(x)= \frac{1}{x} \sum_{n} \int \frac{d^2 \kperpb}{2 (2\pi)^3}  \prod_{i=1}^{n-1} \frac{dy_i}{y_i}\frac{d^2 {{\kappaperpi}}}{2 (2\pi)^3} \, |\Psi_n(\{{y_i,\kappaperpi}\};x,\kperpb,a)|^2 (2\pi)^3 \delta^{(2)}(\kperpb+\sum_{i=1}^{n-1} \kappaperpi) \delta(1-x-\sum_{i=1}^{n-1} y_i)\; ,
\label{eq:pdflf}
\end{multline}
where $\Psi_n(\{{y_i,\kappaperpi}\};x,\kperpb,a)$ is the light-front wave function for $n$ partons. Following \cite{Kovchegov:2012mbw} we shall use the convention where 
$\{{y_i,\kappaperpi}\}$ are $n-1$ spectator partons with longitudinal momentum fractions $y_i$ and
transverse momenta $\kappaperpi$. The density $D_a(x)$ is defined with respect to the parton of type $a$ (where $a$ could be gluon $g$, quark $q$ or antiquark $\bar{q}$) with longitudinal momentum fraction $x$. The   transverse momentum $\kperpb$ of this parton is integrated out and therefore,  the above definition is UV divergent and  needs to be regulated as we shall see below. In the above definition we implicitly assumed the summation over the colors of the outgoing particles as well as their polarizations.  As usual for the light-front calculation, we shall be working in the light cone gauge $A^+=0$. 

\begin{figure}
\begin{center}
\includegraphics[width=0.40\textwidth]{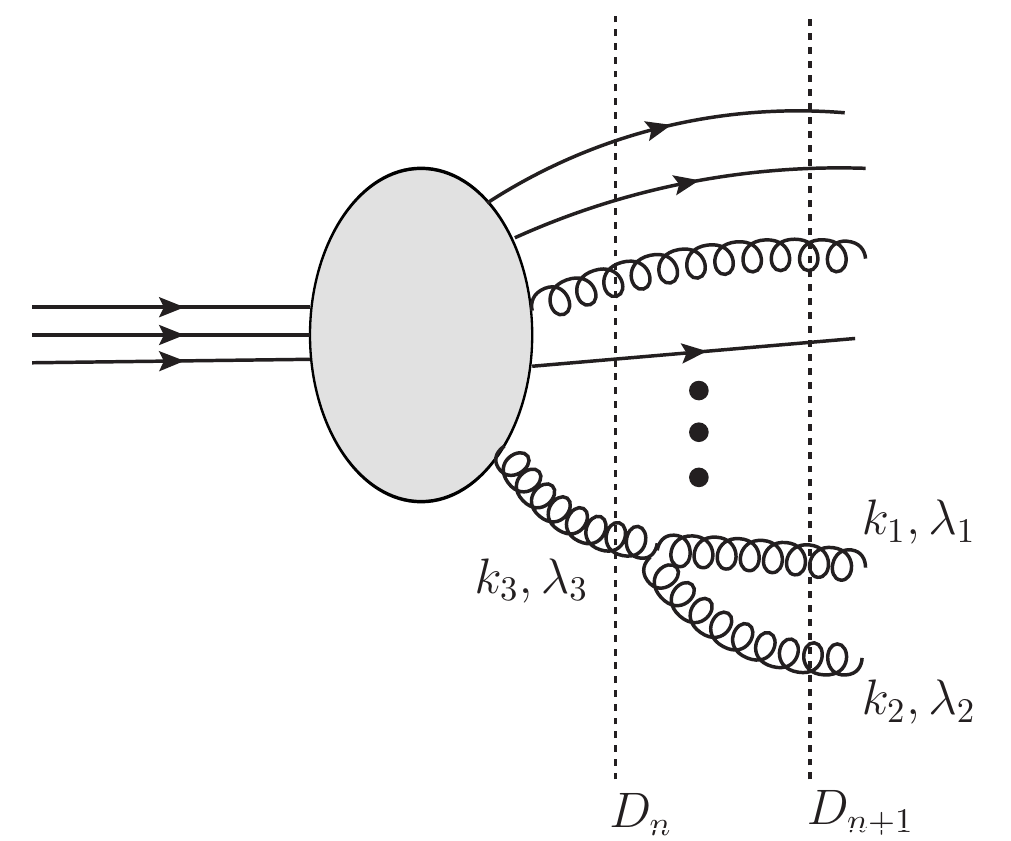}
\caption{Splitting contribution to the proton wave function in the light-front framework. The wave function $\Psi_n$ is  $n$ component wave function of the initial hadron.  The splitting gives contribution to the $\Psi_{n+1}$ wave function. Vertical dashed lines denote light-front energy denominators.
 }\label{fig:lfwf}
\end{center}
\end{figure}

To derive the splitting term contribution to the double parton density, let us focus from now on the gluon-gluon splitting, $g\to gg$; the other channels can be derived in the analogous way.  This contribution is illustrated in Fig.~\ref{fig:lfwf},  where we show the wave function $\Psi_n$ in which one gluon with momentum $(x_3,\kperpthreeb)$ and polarization $\lambda_3$ undergoes the splitting into two daughter gluons with momenta $(x_1,\kperponeb)$ and $(x_2,\kperptwob)$, with the corresponding polarizations $\lambda_1,\lambda_2$. We recall that on the light-front, all of the particles are on-shell and longitudinal  `$+$' and transverse momentum components are conserved while the `$-$'  components are not. Using the rules of the light-front perturbation theory, 
see for example \cite{Cruz-Santiago:2015dla},
we can write  the contribution to the wave function $\Psi_{n+1}$ as
\begin{multline}
\Psi_{n+1}^{AB}(\{{y_i,\kappaperpi}\};x_1,\kperponeb,x_2,\kperptwob) \; = \; V^{\alpha\beta\mu}(k_1,k_2,-k_3)\,
\varepsilon^{\lambda_1 *}_{\alpha }(k_1)\,\varepsilon^{\lambda_2 *}_{\beta }(k_2)\,
\varepsilon^{\lambda_3}_{\mu}(k_3) f^{ABC}
\theta(k_1^+)\, \theta(k_2^+)\,  \frac{1}{D_{n+1}} \\
\times  \frac{1}{k_3^+}\, 
\Psi_n^{C}(\{{y_i,\kappaperpi}\};x_3,\kperpthreeb)  \; ,
\label{eq:phinsplithgg}
\end{multline}
where
$V^{\alpha\beta\mu}(k_1,k_2,-k_3) $ is the triple gluon vertex (with all momenta outgoing) and  $\varepsilon^{\lambda}_{\mu}(k)$ are polarization vectors defined as
\be
\varepsilon^{\lambda}_{\mu} (k) = (0,\frac{2 \vec{\varepsilon}_{\perp}^{\,\lambda} \cdot \kperpb}{\eta\cdot k},\vec{\varepsilon}_{\perp}^{\,\lambda}) \; . 
\ee
In the above definition,  the light-like vector $\eta=(0,1,0,0)$ and the two-dimensional  transverse polarization vectors are defined as 
$\vec{\varepsilon}_{\perp}^{\,\pm} =\frac{1}{\sqrt{2}}(\pm 1,i)$. In the notation of the vectors we have  used standard convention on the light-front where $k^{\mu}=(k^+,k^-,\kperpb)$ with $k^{\pm}=k^0\pm k^3$ and
$\kperpb=(k^1,k^2)$, thus $k^2=k^+ k^--\kperpb^2$.  The light-front denominator $D_{n+1}$ is defined  as the difference between the light-front energies for the intermediate state and the initial state
\be
D_{n+1} = k_1^-+k_2^- + \sum_{i=1}^{n-1} \kappa_i^- - P^- \;,
\ee
where $P^-$ is the light-front energy of the incoming hadron.  In Eq.~\eqref{eq:phinsplithgg} we also reinstated  explicit colors of the gluons. We also introduce the following  notation from 
\cite{Motyka:2009gi}
\be
v_{ij} = \vec{\varepsilon}_{\perp}^{\,+} \cdot\left( \frac{ \mathbf{k}_{i\perp}}{x_i}-\frac{\mathbf{k}_{j\perp}}{x_j}\right),~~~~~~~~~~~~~~~~~~~~
 v^*_{ij} = \vec{\varepsilon}_{\perp}^{\,-} \cdot\left( \frac{ \mathbf{k}_{i\perp}}{x_i}-\frac{\mathbf{k}_{j\perp}}{x_j}\right) ,
\ee
where $x_i = {k_i^+}/{P^+}$.
It is easy to see that the contraction of the triple gluon vertex with the polarization vectors leads to the following expressions (see Table 2  in \cite{Cruz-Santiago:2015dla})
\be
V^{+\rightarrow ++}  = 2ig x_3\, v_{21},~~~~~~V^{+\rightarrow +-} = 2ig x_1\, v^*_{32},
~~~~~V^{+\rightarrow -+} = 2ig x_2\, v^*_{13} \;.
\label{eq:vertices}
\ee
Here, $k_3$ is incoming and $k_1,k_2$ are outgoing momenta, i.e. $k_3\rightarrow k_1 k_2$.
Let us first see how the splitting term in the collinear kinematics arises from these expressions. In the collinear case, one assumes strong ordering of the transverse momenta. Hence,
$\kperponeb\simeq -\kperptwob \simeq \kperpb$ and 
$\kperp \gg \kperpthree$.  In this approximation the light-front denominator is given by
\be
D_{n+1} \; \simeq \; \frac{\kperp^2 x_3}{x_1 x_2}\frac{1}{P^+}\;,
\ee
and the vertices in Eq.~(\ref{eq:vertices}) reduce to
\be
V^{+\rightarrow ++}  = -2ig x_3^2\,\frac{\vec{\varepsilon}_{\perp}^{\,+} \cdot \kperpb}{x_1 x_2},
~~~~~V^{+\rightarrow +-}  = 2ig x_1\, \frac{\vec{\varepsilon}_{\perp}^{\,-} \cdot \kperpb}{x_2},
~~~~~V^{+\rightarrow -+} = 2ig x_2\, \frac{\vec{\varepsilon}_{\perp}^{\,-} \cdot \kperpb}{x_1}  \;.
\label{eq:vertices}
\ee
Squaring the wave function, summing over the final state polarizations and colors, one obtains, see
also \cite{Kovchegov:2012mbw},
\be
|\Psi_{n+1}(\{{y_i,\kappaperpi}\};x_1,\kperponeb;x_2,\kperptwob)|^2 = 8\pi \alpha_s\,  \frac{x_1 x_2}{\kperp^2 x_3^2} \,P_{gg}^{(0)}\,\Big(\frac{x_1}{x_3}\Big) 
|\Psi_n(\{{y_i,\kappaperpi}\};x_3,\kperpthreeb)|^2\; .
\ee
Using this result one can re-derive the standard DGLAP evolution equation for the single parton distribution function as demonstrated in \cite{Kovchegov:2012mbw}. However, we are interested in the contribution to the double parton distribution function. Therefore, after
integrating the wave function  $\Psi_{n+1}$ over the transverse momenta $\kperponeb$ and $\kperptwob$ and over the spectator momenta, but keeping both the longitudinal momentum fractions $x_1$ and $x_2$ fixed, we obtain the following contribution to the non-homogeneous part of the double integrated distribution function:
\begin{multline}
D_{gg}^{nh}(x_1,x_2)\;=\;\frac{1}{x_1 x_2} \sum_{n} \int \frac{d^2 \kperponeb}{2 (2\pi)^3} \int \frac{d^2 \kperptwob}{2 (2\pi)^3} 
\prod_{i=1}^{n-1} \frac{dy_i}{y_i}\frac{d^2 {\kappaperpi}}{2 (2\pi)^3}
\\
\times |\Psi_{n+1}(\{{y_i,\kappaperpi}\};x_1,\kperponeb,x_2,\kperptwob)|^2 (2\pi)^3\, \delta^{(2)}(\kperponeb+\kperptwob+\sum_{i=1}^{n-1} \kappaperpi)\,\delta(1-x_1-x_2-\sum_{i=1}^{n-1} y_i) =\\
\frac{1}{x_1 x_2} \int \frac{d^2 \kperpb}{2 (2\pi)^3} 8\pi \alpha_s  \frac{x_1 x_2}{\kperp^2 x_3} 
P_{gg}^{(0)}\Big(\frac{x_1}{x_3}\Big)  \frac{1}{x_3}
\int \frac{d^2 \kperpthreeb}{2 (2\pi)^3}  \sum_{n}  
\prod_{i=1}^{n-1} \frac{dy_i}{y_i}\frac{d^2 {\kappaperpi}}
{2 (2\pi)^3} \,  |\Psi_n(\{{y_i,\kappaperpi}\};x_3,\kperpthreeb)|^2 
\\
\times(2\pi)^3 
\delta^{(2)}(\kperpthreeb+\sum_{i=1}^{n-1} \kappaperpi)\, \delta(1-x_3-\sum_{i=1}^{n-1} y_i)\,,
   \label{eq:dpdflf} 
\end{multline}
where we changed the integrated variables from $\kperponeb,\kperptwob$ to the sum and the difference i.e. $\kperpthreeb=\kperponeb+\kperptwob$ and $\kperpb=\frac{1}{2}(\kperponeb-\kperptwob)$. There are two integrals over the transverse momenta $\kperpthreeb$  and $\kperpb$   that need to be regulated. Since we are working in the collinear regime ($\kperp \gg \kperpthree$) then we have that  the $\kperpthree$ integral is regulated by the $\kperp$ and it gives the integrated parton density
\begin{equation}
  \frac{1}{x_1 x_2}  \int \frac{d^2 \kperpb}{2 (2\pi)^3}\, 8\pi \alpha_s\,  \frac{x_1 x_2}{\kperp^2 x_3}\, P_{gg}^{(0)}\Big(\frac{x_1}{x_3}\Big) D_g(x_3,\kperp) \;.
  \end{equation}
The integral over $\kperpb$ in the above needs to be regulated as well with a UV cutoff. Introducing the  scale $\mu$, we finally obtain the splitting contribution to the double parton distribution function
\begin{equation}
 \frac{\alpha_s}{2\pi} \int^{\mu^2} \frac{d\kperp^2}{\kperp^2} \frac{1}{x_3} P_{gg}^{(0)}\Big(\frac{x_1}{x_3}\Big) D_g(x_3,\kperp) =  \frac{\alpha_s}{2\pi}  \frac{1}{x_1+x_2} P_{gg}^{(0)}\Big(\frac{x_1}{x_2+x_1}\Big) \int^{\mu^2} \frac{d\kperp^2}{\kperp^2}  D_g(x_1+x_2,\kperp) \; ,
\end{equation}
where we used the fact that $x_3=x_1+x_2$.
The left hand side of Eq.~\eqref{eq:dpdflf} can be interpreted as the integrated double parton density $D_{gg}(x_1,x_2)$ and thus the last equation gives the splitting contribution to the DPDF. The other channels can be obtained similarly.
Differentiating with respect to $\mu$ we have that this gives the contribution  to the inhomogeneous part of the evolution equation
\begin{equation}
\frac{\partial }{\partial \ln {\mu^2}}D_{a_1 a_2}(x_1,x_2,\mu,\mu) = \frac{\alpha_s}{2\pi}\, \sum_{a}
P_{a_1 a}^{(0)}\Big(\frac{x_1}{x_1+x_2}\Big)\frac{D_a(x_1+x_2,\mu)}{x_1+x_2}  \; ,
\label{eq:dpdfnh2}
\end{equation}
where we reinstated flavor indices $a,a_1,a_2$ to include other  parton transitions. The above equation is equivalent to the non-homogeneous part of Eq.~\eqref{eq:24}, together with Eq.~\eqref{eq:10}.

\subsection{Transverse momentum dependence in parton splitting}

Let us now see, how the transverse momentum dependence can be introduced into the splitting. To this aim, we can go back to  Eq.~\eqref{eq:dpdflf}  and analyze the integrand of this expression. In addition we need to keep track of the transverse momenta after the splitting while still working in the strong ordering approximation. Thus we shall assume, $\kperp\simeq \kperpone \simeq \kperptwo  \gg \kperpthree$.
We obtain (for the gluon-gluon splitting case)\footnote{In Eq.~\eqref{eq:unpdf1} there is an angular dependence in the transverse momenta $\kperponeb,\kperptwob$. The unintegrated function with such a dependence  is therefore integrated in the cross section with the measure $\frac{d^2 \kperponeb}{\pi \kperpone^2}\frac{d^2 \kperptwob}{\pi \kperptwo^2}$. In the case when there is no angular dependence in the unintegrated function, like in Eq.~\eqref{eq:unpdf1}, the measure needs to be taken as $\frac{d \kperpone^2}{\kperpone^2} \frac{d \kperptwo^2}{ \kperptwo^2}$.}
\begin{equation}
f_{a_1 a_2}(x_1,x_2,\kperponeb,\kperptwob) =  \, \frac{\alpha_s}{2\pi}  \frac{1}{x_1+x_2}\,
\frac{\kperpone^2 \kperptwo^2}{\kperpthree^2 \kperp^2}\, P_{a_1 a}^{(0)}\Big(\frac{x_1}{x_2+x_1}\Big)\,   
f_a(x_1+x_2,\kperpthreeb) \; .
\label{eq:unpdf1}
\end{equation}
This contribution has been previously derived 
 in \cite{Diehl:2011yj}, with some modifications which include the polarizations of the produced partons. In that case there  is  also a second parton distribution, the Boer-Mulders function \cite{Mulders:2000sh,Boer:1997nt} which describes polarized partons in the unpolarized hadron. The contribution from it vanishes when the angular integrals are performed.   

For practical applications we propose to   utilize the  formula (\ref{eq:unpdf1})  with the unintegrated PDFs modeled according to the KMR approach,  discussed in Sec.~\ref{sec:skmr},
\be
\label{eq:spdfrepeat}
f_a(x_1+x_2,\kperpthree,Q) = T_a(Q,\kperpthree) \sum_{a'} \int_{x_1+x_2}^{1-\Delta} \frac{dz}{z} P_{aa'}(z,\kperpthree) \,
D_{a'}\!\!\left(\frac{x_1+x_2}{z},\kperpthree\right).
\ee
In such a case, the unintegrated double distributions (\ref{eq:unpdf1}) become scale dependent with equal scales, $Q_1=Q_2=Q$,
\begin{equation}
f_{a_1 a_2}(x_1,x_2,\kperpone,\kperptwo,Q,Q) =  \, \frac{\alpha_s}{2\pi}  \frac{1}{x_1+x_2} \frac{\kperpone^2 \kperptwo^2}{\kperpthree^2 \kperp^2} P_{a_1 a}^{(0)} \Big(\frac{x_1}{x_2+x_1}\Big)   f_a(x_1+x_2,\kperpthree,Q) \; .
\label{eq:unpdf2}
\end{equation}
The  reason is that this formula only contains evolution of the unintegrated parton density 
up to a scale $Q$ and then the splitting is treated with the  transverse momentum dependence.
The two partons from the splitting should evolve now. However, the initial partons have nonzero
transverse momenta which may be from the perturbative region, $\kperpone, \kperptwo \ge Q_0$. Thus, we should consider 
QCD radiation  with transverse momentum dependent splitting functions, see e.g. \cite{Hautmann:2012sh,Hentschinski:2016wya}. 
We postpone considering such a case for a future publication with a numerical analysis.

Several important comments are in order here. First of all, the scale $Q$ which appears on the right hand side  in Eq.~\eqref{eq:unpdf2} is the scale  which can be related to the cutoff on the transverse momentum $\kperpthree$. The assumption that we are making is that it is the same scale which appears on the left hand side of Eq.~\eqref{eq:unpdf2} for the double parton distribution function. This is motivated by the structure  in the integrated form of the non-homogeneous part of Eq.~\eqref{eq:dpdfnh2}. That is Eq.~\eqref{eq:unpdf2}, when integrated over the transverse momentum, leads to Eq.~\eqref{eq:dpdfnh2}. 

 Second, in the derivation of the splitting it is possible to include the transverse momentum transfer $\rperpb$. The more general formula (still in the approximation of the  leading power), which includes the dependence on this variable,  has been derived in \cite{Diehl:2011yj},
 \be
f_{a_1 a_2}(x_1,x_2,\kperponeb,\kperptwob,\rperpb;Q,Q) = \, \frac{\alpha_s}{2\pi} \, \frac{\kperpone^2 \kperptwo^2}{\kappaperp^2} \frac{(\kperpb+\frac{1}{2}\rperpb) \cdot (\kperpb-\frac{1}{2}\rperpb)}
{(\kperpb+\frac{1}{2}\rperpb)^2 (\kperpb-\frac{1}{2}\rperpb)^2} 
 \frac{f_a(x_1+x_2,\kappaperp,Q)}{x_1+x_2} \,\, P^{(0)}_{a_1 a}  \!\!\left(\frac{x_1}{x_1+x_2}\right),
\label{eq:nh2}
\ee
which clearly reduces to the previous formula when $\rperpb\rightarrow 0$. As has been discussed in \cite{Blok:2010ge}, this dependence should be taken into account through the appropriate form factor.  Finally, it is possible to go beyond the leading power approximation and derive splittings with more exact kinematics with transverse momentum dependence. We shall leave the systematic analysis of these  improvements to the future work \cite{kgbasinprogress}.

\section{Conclusions}

We presented a construction of the unintegrated double parton distribution functions which depend on parton transverse momenta, $\kperpone$ and 
$\kperptwo$, in addition to their two longitudinal momentum fractions, $x_1$ and $x_2$, and two factorization scales, $Q_1$ and $Q_2$.
We  follow the method proposed by Kimber, Martin and Ryskin in \cite{Kimber:1999xc} to construct unintegrated 
single parton distribution functions, which relies on unfolding the last step in the DGLAP evolution of the  integrated PDFs. We found two contributions to the unintegrated DPDFs, corresponding to  the possibility that the two partons originate either from the proton or from the splitting of a single parton. In the first case, the unintegrated DPDFs are given by Eqs.~(\ref{eq:d23a})-(\ref{eq:d23c}), presented in the $x$-space.
They correspond to four regions of transverse momenta, illustrated in Fig.~\ref{fig:fig2}. 

The perturbative case with parton splitting is more involved. We analyzed two cases, the unfolding of the transverse momentum dependence from the last step in the DGLAP evolution of two partons, and the case where transverse momenta are generated directly from  the single parton  splitting  into two partons.
In the first case, we found that only formula  (\ref{eq:c6}) is acceptable for the unintegrated DPDFs from the point of view of the $\kperp$-factorization of the double parton scattering cross sections. In the second case, we propose formula \eqref{eq:unpdf2}, 
which includes transverse momentum dependence generated from the perturbative splitting of one parton into two daughter partons. In that case, the KMR prescription is applied to the single PDF, in order to introduce the transverse momentum dependence, and then the splitting is treated by including the transverse momentum dependence. We kept the derivation in the strong ordering approximation to be consistent with the rest of the framework.

It also should be mentioned that  in our discussion we neglected the spin  dependence by considering the spin averaged case.  We also do not  discuss the dependence on the additional transverse momentum, the momentum transfer  
$\rperpb$, setting it to zero. In practical applications such a dependence is modeled with an appropriate form factor, see \cite{Blok:2010ge,Golec-Biernat:2014nsa,Gaunt:2014rua}. The  framework for the modeling of the transverse momentum dependence proposed in these papers could be now implemented into a numerical code. The  detailed numerical analysis of the unintegrated DPDFs formulated in this paper will be presented elsewhere  \cite{kgbasinprogress}.

\begin{acknowledgments}
This work was supported by the National Science Center, Poland, Grant No. 2015/17/B/ST2/01838,   
by the Department of Energy  Grant No. DE-SC-0002145 and by the Center
for Innovation and Transfer of Natural Sciences and Engineering Knowledge in Rzesz\'ow.
\end{acknowledgments}

\appendix

\label{app:1}
\section{Proof}

We will prove that Eq.~(\ref{eq:19}) with equal scales, $Q_1=Q_2\equiv Q$, 
is the solution to the equation (\ref{eq:24}) which is the LO  evolution equation 
for the DPDFs. This  means that the splitting functions in this equation
are proportional to the LO strong coupling constant, $\alpha_s(Q)/(2\pi)$, which can be absorbed in a new evolution
parameter
\be
\label{eq:aa0}
t=\frac{6}{33-2n_f}\ln\frac{\ln(Q^2/\Lambda^2_{QCD})}{\ln(Q^2_0/\Lambda^2_{QCD})}\,,
\ee
where $n_f$ is the number of active quark flavors.
Thus, Eq.~(\ref{eq:24})  takes the the following form 
\be\nonumber
\label{eq:aa3}
\frac{\partial}{\partial t} \Dtilde_{a_1a_2}(n_1,n_2,t)\eq \sum_{\aprime}
 \Ptilde_{a_1a\prime}(n_1) 
\Dtilde_{\aprime a_2}(n_1,n_2,t)
- S_{a_1}\Dtilde_{a_1a_2}(n_1,n_2,t)
\\\nonumber
\\
&+&\sum_{\aprime} \Ptilde_{a_2\aprime}(n_2) \Dtilde_{a_1\aprime}(n_1,n_2,t)
-S_{a_2}\Dtilde_{a_1a_2}(n_1,n_2,t) 
\,+\,
\Dtilde_{a_1a_2}^{(sp)}(n_1,n_2,t),
\ee
where $ \Ptilde_{a\aprime}(n)$ are the LO splitting functions in the Mellin space (anomalous dimensions),  
$\Dtilde_{a_1 a_2}^{(sp)}$ is given by Eq.~(\ref{eq:19b}), and we defined
\be
\label{eq:aa5}
S_{a} = \sum_{\aprime} \int_0^{1} dz z \Phat_{\aprime a}(z)\,.
\ee
Eq.~(\ref{eq:aa3}) can be written in the matrix form with respect to flavor indices,
\be
\label{eq:aa6}
\frac{\partial}{\partial t} \Dtilde(n_1,n_2,t) = {\cal P}(n_1)\,\Dtilde(n_1,n_2,t) +\Dtilde(n_1,n_2,t)\, {\cal P}^\dagger(n_2)+ \Dtilde^{(sp)}(n_1,n_2,t)\,,
\ee
where the matrices $\Dtilde= (\Dtilde_{ab}),\,\Dtilde^{(sp)}= (\Dtilde^{(sp)}_{ab})$ and 
${\cal{P}}=(\Ptilde_{ab}-S_{a}\,\delta_{ab})$.
To solve Eq.~(\ref{eq:aa6}), we postulate the solution in the form
\be
\label{eq:aa9}
\Dtilde(n_1,n_2,t) = \Etilde(n_1,t)\,\Dtilde^0(n_1,n_2,t)\,\Etilde^\dagger(n_2,t).
\ee
Eq.~(\ref{eq:aa6}) is fulfilled if
\be
\label{eq:aa10}
\frac{d\Etilde(n,t)}{dt} = {\cal P}(n)\,\Etilde(n,t)\; ,
\ee
with the initial condition $\Etilde(n,0)=1$, and 
\be
\label{eq:aa11}
\Ehat(n_1,t)\frac{d\Dhat^0(n_1,n_2,t)}{dt}\Ehat^\dagger(n_2,t)=\Dtilde^{sp}(n_1,n_2,t)\,.
\ee
Notice that Eq.~(\ref{eq:aa10}) is equivalent to Eq.~(\ref{eq:b15}) after changing the evolution variable to $t$. Thus, $\Ehat(n,t)$
is the parton-to-parton  evolution distribution introduced in Section~{\ref{sec:ad}}.
The solution to Eq.~(\ref{eq:aa10}) reads 
\be
\label{eq:a18c}
\Etilde(n)={\rm e}^{{\cal{P}}(n)t}\,,
\ee
therefore, it fulfills the relation
\be
\label{eq:aa12}
\Etilde(n,t_1)\Etilde(n,t_2)=\Etilde(n,t_1+t_2)\,,
\ee
which can be used to write Eq.~(\ref{eq:aa11})  in the form
\be
\label{eq:aa13}
\frac{d\Dtilde^0(n_1,n_2,t)}{dt}=
\Etilde(n_1,-t)\,\Dtilde^{(sp)}(n_1,n_2,t)\,\Etilde^\dagger(n_2,-t)\,.
\ee
 Its solution is given by
\be
\label{eq:aa14}
\Dtilde^0(n_1,n_2,t)=\Dtilde(n_1,n_2,0)+
\int_0^tdt^\prime
\Etilde(n_1,-t^\prime)\,
\Dtilde^{(sp)}(n_1,n_2,t^\prime)\,\Etilde^\dagger(n_2,-t^\prime) \,,
\ee
where $\Dtilde(n_1,n_2,0)$ is an initial condition.
Substituting (\ref{eq:aa14}) into (\ref{eq:aa9}), we  find the final form of the solution,
\be
\label{eq:aa15}
\Dtilde(n_1,n_2,t)\eq  \Etilde(n_1,t)\,\Dtilde(n_1,n_2,0)\,\Etilde^\dagger(n_2,t)
+\int_0^t d \tprime\, \Etilde(n_1,t-\tprime)\,\Dtilde^{(sp)}(n_1,n_2,\tprime)\,\Etilde^\dagger(n_2,t-\tprime)\,,
\ee
which is equivalent to relation (\ref{eq:19}).

\bibliographystyle{h-physrev4}
\bibliography{mybib}

\end{document}